\documentclass[aps,twocolumn,showpacs,preprintnumbers,nofootinbib,prd,superscriptaddress,groupedaddress,10pt]{revtex4-1}

\makeatletter
\def\l@subsubsection#1#2{}
\def\l@subsubsub section#1#2{}
\makeatother

\usepackage{amsmath}
\usepackage{graphicx,amssymb,amsmath,amsthm,amsfonts,epsfig,epsf}
\usepackage[usenames,dvipsnames,svgnames,table]{xcolor}
\usepackage{epstopdf}
\definecolor{darkred}{rgb}{0.5,0,0}
\usepackage{aas_macros}
\usepackage{bm}
\usepackage{dcolumn}
\usepackage[utf8]{inputenc}
\usepackage{latexsym}
\usepackage{rotating}
\usepackage{longtable}

\usepackage{enumerate}
\usepackage{tensor}
\usepackage{float}
\usepackage{multirow}
\usepackage{booktabs}
\usepackage{mathtools}
\usepackage{url}
\usepackage[linktocpage,hidelinks]{hyperref}

\newcommand{\til}{~}

\usepackage{colortbl}

\def\nn{\nonumber}

\def\be{\begin{equation}}
\def\ee{\end{equation}}
\newcommand{\beq}{\begin{eqnarray}}
\newcommand{\eeq}{\end{eqnarray}}

\newcommand{\nocontentsline}[3]{}
\newcommand{\tocless}[2]{\bgroup\let\addcontentsline=\nocontentsline#1{#2}\egroup}
\def\ba{\begin{align}}
\def\ea{\end{align}}

\newcommand{\warn}[1]{{\textcolor{red}{\sf{[IN PROGRESS]}} }}

\begin{document}
\title{Non-linear tides and Gauss-Bonnet scalarization}

\author{
Lorenzo Annulli$^{1}$ and  %
Carlos A. R. Herdeiro$^{1}$
}

\affiliation{${^1}$ Departamento de Matem\'{a}tica da Universidade de Aveiro and Centre for Research and Development in Mathematics and Applications (CIDMA), Campus de Santiago, 3810-183 Aveiro, Portugal}

\begin{abstract}
In linear perturbation theory, a static perturber in the vicinity of a Schwarzschild black hole (BH) enhances [suppresses] the Gauss-Bonnet (GB) curvature invariant, $\mathcal{R}_{\rm GB}$, in the high [low] tide regions. By analysing exact solutions of the vacuum Einstein field equations describing one or two BHs immersed in a multipolar gravitational field, which is locally free of pathologies, including conical singularities, we study the corresponding \textit{non-linear tides} on a fiducial BH, in full General Relativity (GR). We show that the tidal field due to a far away, or close by, static BH creates high/low tides that can deviate not only quantitatively but also qualitatively from the weak field/Newtonian pattern. Remarkably, the suppression in low tide regions never makes $\mathcal{R}_{\rm GB}$ negative on the BH, even though the horizon Gaussian curvature may become negative; but $\mathcal{R}_{\rm GB}$ can vanish  in a measure zero set, a feature qualitatively recovered in a Newtonian analogue model. Thus, purely gravitational, static, tidal interactions in GR, no matter how strong, cannot induce GB$^-$ scalarization. We also show that a close by BH produces noticeable asymmetric tides on another (fiducial) BH.

\end{abstract}

\maketitle

\section{Introduction}
In his 1687 \textit{Principia Mathematica}~\cite{Newton:1687eqk}, Newton brilliantly understood that tides, and their mysterious occurrence \textit{twice} a day, are explained by his law of universal gravitation. Newton used a \textit{static}, simplified model, to extract the main effect. The Earth, covered by a uniforme  deformable fluid layer (total mass $M$, total radius $R$), is under the gravitational action of a point-like companion (the Moon, mass $M_c$), placed at $\theta_c=0$ and distance $r_c$, on an Earth centred polar-frame. Then, the difference, $\Delta {\bf F}$, between the Moon's force on a peripherical point $(R,\theta)$  and on  Earth's centre ($r=0$) is: directed away from Earth's centre  at $\theta=0,\pi$; directed towards Earth's centre  at $2\theta=\pi,3\pi$. Generically, the magnitude is, to leading order in $R/r_c$ (using, throughout, $G=1=c$),
\begin{equation}
\frac{|\Delta{\bf  F}|}{M R}\simeq \frac{M_c}{r_c^3}\sqrt{1+3\cos^2\theta} \ .
\label{newtontides}
\end{equation}
In essence, Newton showed that differential gravitational forces  on an extended object, due to a perturber, are \textit{quadrupolar}, which explains the bidaily tides, with two symmetric ($i.e.$ equal magnitude, in this approximation) high tides (HTs) twice as strong as low tides (LTs). As a tribute to their most notable physical effect, differential gravitational forces became known as \textit{tidal forces}.

In General Relativity (GR)~\cite{Einstein:1915ca} tidal forces acquire an even more fundamental significance. The gravitational force is non-tensorial and is (locally) gauged away in a freely falling frame, as a consequence of the equivalence principle. Only tidal  forces are covariant, tensorially described by the Riemann tensor, $R^\mu_{\ \nu\alpha\beta}$, encoding them as the spacetime curvature.  In other words,  tidal forces are the invariant signature of gravity in GR. Despite the conceptual discontinuity, however, tidal effects are similar to their Newtonian counterparts in appropriate situations, and not exclusively in the weak field limit. 

Consider an infinitesimal sized collection of test particles freeling falling into a Schwarzschild black hole (BH) with ADM  mass  $M$ and described in standard Schwarzschild coordinates $(t,r,\theta,\varphi)$ by the metric $g_{\mu\nu}^{\rm Sch}$. The relative acceleration of two of these particles connected by the spacelike vector $\eta^\alpha$ is given by the geodesic deviation equation~\cite{Wald:1984rg}. Working with the standard Schwarzschild orthonormal  frame $e^\mu_{\ \hat{\mu}}$ (see $e.g.$~\cite{dInverno:1992gxs,Lima:2020wcb}), the radial and angular tidal accelerations on the test particles is
\begin{equation}
\frac{D^2\eta^{\hat{r}}}{D \tau^2}=\frac{2M}{r^3}\eta^{\hat{r}} \ , \quad \frac{D^2\eta^{\hat{\theta}}}{D \tau^2}=-\frac{M}{r^3}\eta^{\hat{\theta}} \ .
\label{testtides}
\end{equation}
Thus, not only one recovers the quadrupolar effect, but the quantitative acceleration on the radial and angular direction exactly matches that of the Newtonian case~\eqref{newtontides}, with the ``HT" twice as strong as the ``LT".

\begin{figure}
\centering
\includegraphics[width=6.5cm,keepaspectratio]{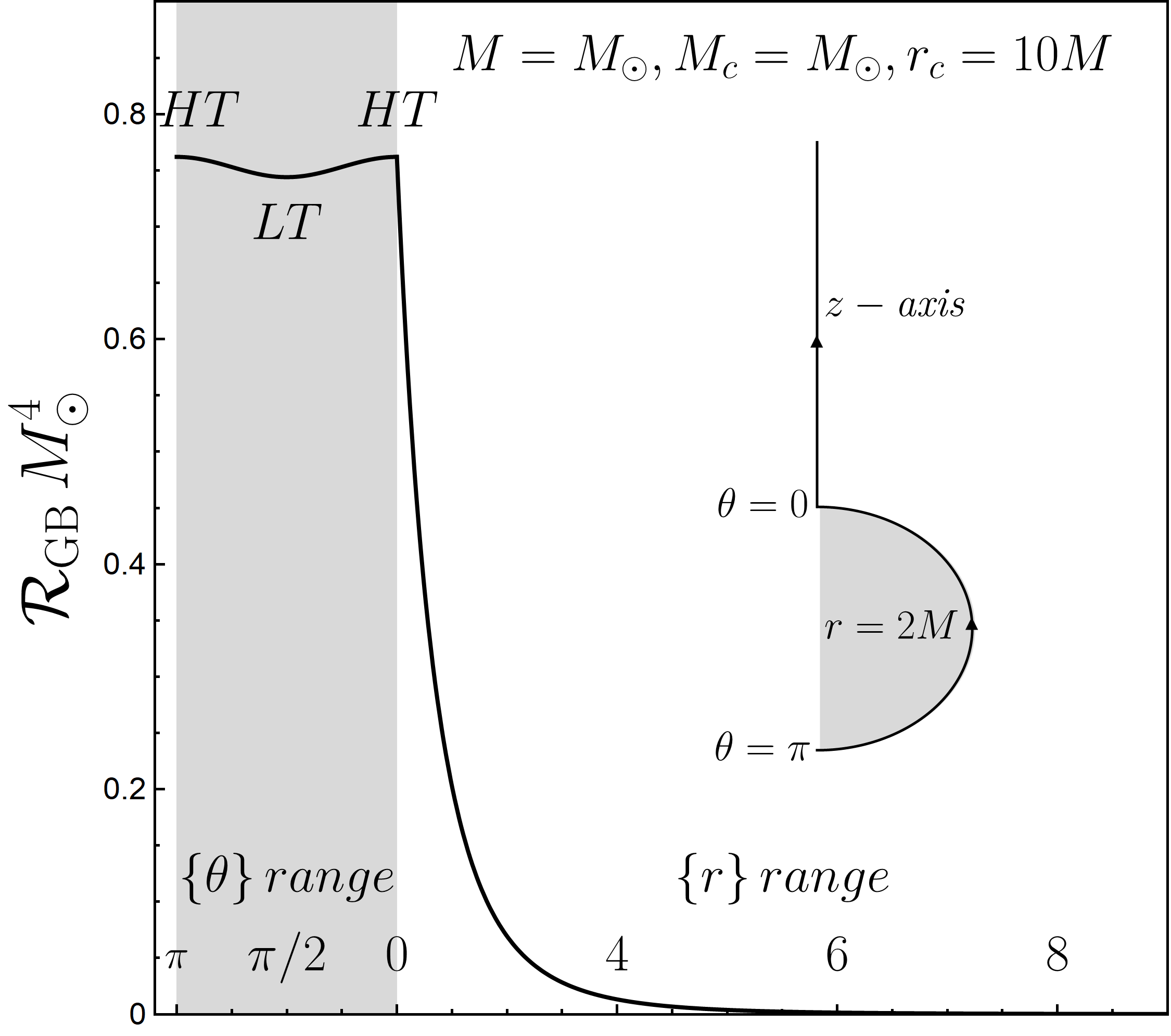} 
\caption{GB invariant along the $z$-axis (outside the horizon) and on the horizon of a Schwarzschild BH immersed in the quadrupolar tidal field caused by a distant companion, $cf.$ Eq.\til\eqref{eq:GB_perturb}. The line followed in physical space is illustrated on the right side and mimics a $z$=const. line in Weyl coordinates.}
\label{fig:plotRGB_perturbation}
\end{figure}

The result~\eqref{testtides} holds for test particles. In reality, additional masses tidally distort the Schwarzschild BH. This can be quantitatively addressed in perturbation theory. Consider now a  perturber companion with mass $M_c$ in the BH spacetime. Assume that the pertuber is placed at $(r_c, \theta_c=0, \phi_c=0)$ and, like Newton, that it remains static in the sky of the central Schwarzschild BH. Then, metric perturbations are split into  axial and polar~\cite{Regge:1957td}, with the former being subleading~\cite{Poisson:2005pi,Cardoso:2021qqu}. Focusing on the latter, the leading effect is quadrupolar, and at linear order the perturbed line element reads, 
\begin{equation}
ds^2=g_{\mu\nu}^{\rm Sch}+\epsilon h_{\mu\nu} \ , \qquad  \epsilon\equiv  M^2 \frac{ M_c }{ r_c^3} \ ,
\label{perturbedm}
\end{equation}
with 
\begin{align}\label{eq:metric_func_schw_tidal}
&h_{tt}=\frac{8 \pi }{5M^2} r^2 f^2  \sum_m Y^{2m}\left(\theta, \phi\right) Y^{2m} \left(0, 0\right)\nn\,,\\
&h_{rr}=f^{-2} h_{tt}\nn\,,\\
&h_{\theta\theta}=\frac{8 \pi }{5M^2} r^2 (r^2-2M^2) \sum_m Y^{2m}\left(\theta, \phi\right) Y^{2m} \left(0, 0\right)\nn\,,\\
&h_{\phi\phi}= h_{\theta\theta}\sin^2\theta \, \,,
\end{align}
where $f\equiv 1-2M/r$ and $Y^{lm}(\theta,\phi)$ are the standard spin-0 spherical harmonics. To assess the leading tidal effect it is simpler to look at a scalar curvature invariant, for which the natural choice in vacuum is the Kretschmann scalar $R_{\mu\nu\alpha\beta}R^{\mu\nu\alpha\beta}$ (which is our case, modulo the perturber). Since, in vacuum, this invariant coincides with the Gauss-Bonnet (GB) invariant,
\begin{align}
\mathcal{R}_{\rm GB}\equiv R_{\mu\nu\alpha\beta}R^{\mu\nu\alpha\beta}-4 R_{\mu\nu}R^{\mu\nu}+R^2 \  ,
\label{GB}
\end{align}
where $R_{\mu\nu}$ and $R$ are the Ricci tensor and scalar, respectively, we shall focus the discussion instead on the GB invariant, having in mind a practical application, which we will explore below. However, it is important to note that the concepts and principles discussed here can be equally applied to the Kretschmann invariant. For the perturbed metric~\eqref{perturbedm}-\eqref{eq:metric_func_schw_tidal} the GB invariant reads, 
\begin{align}\label{eq:GB_perturb}
\mathcal{R}_{\rm GB}= 48 \left(\frac{M}{ r^3}\right)^2+  24[1+3 \cos (2 \theta )] \frac{M}{ r^3} \frac{ M_c }{ r_c^3}\,,
\end{align}
where the first term is the unperturbed Schwarzschild BH GB invariant. One observes that the tidal quadrupolar effect increases symmetrically the  HT scalar invariant, occuring along the line of sight of the companion $(\theta=0,\pi)$, and decreases the corresponding LT invariant (by half the amount) on the orthogonal direction  $(2\theta=\pi,3\pi)$. In this sense, the tides on the BH still mimic the Newtonian tides. 
To illustrate the tides just described in perturbation theory, and setting the stage for the analysis below where Weyl coordinates~\cite{Stephani:2003tm} will be used, Fig.\til\ref{fig:plotRGB_perturbation} exhibits the value of $\mathcal{R}_{\rm GB}$ in  Eq.\til\eqref{eq:GB_perturb}  (black solid curve) \textit{both}  outside the horizon - keeping along the $z$-axis - \textit{and} along the horizon - fixing $r=2M$, while $\theta$ spans $]0,\pi[$. The line followed in physical space is illustrated in the right hand side of the plot, with the BH region shaded in grey in both sides of the figure.

At this point let us introduce two quantitative measures of (possibly non-linear) tides on a fiducial BH under some external action, described in terms of curvature invariants. The first of these measures describes the HT/LT \textit{differential ratio}; we introduce this quantity for both the close (to the perturber) and the far HT regions as
\begin{equation}
\Delta T^{\rm close}\equiv \frac{\Delta  \mathcal{R}_{\rm GB} (\theta=0)}{\Delta  \mathcal{R}_{\rm GB} (\theta=\pi/2)} \ , \Delta T^{\rm far}\equiv \frac{\Delta  \mathcal{R}_{\rm GB} (\theta=\pi)}{\Delta  \mathcal{R}_{\rm GB} (\theta=\pi/2)} \ ,
\label{deltat}
\end{equation}
where $\Delta  \mathcal{R}_{\rm GB}$ is obtained by subtracting from the total $\mathcal{R}_{\rm GB}$ the contribution from the unperturbed BH, which in the case of~\eqref{eq:GB_perturb} is simply to drop the first term. Obviously in the case of~\eqref{eq:GB_perturb}, 
\begin{equation}
\Delta T^{\rm close}=-2=\Delta T^{\rm far} \ ,
\end{equation}
which coincides with the tidal differential ratio seen also in the Newtonian case and informs us the tides are symmetric. The measure in Eq.~\eqref{deltat} is quite natural and informative, but it requires a dictionary between the perturbed and unperturbed cases in terms of the same physical quantities. We will illustrate below that this can be challenging.

Another quantity describing the (possibly non-linear) tides is simply the HT/LT  \textit{ratio}, that compares HTs and LTs directly, without the need of introducing the unperturbed  background, avoiding the aforementioned difficulty. The HT/LT  ratio is defined as
\begin{equation}
\delta T^{\rm close}\equiv \frac{ \mathcal{R}_{\rm GB} (\theta=0)}{ \mathcal{R}_{\rm GB} (\theta=\pi/2)} \ , \delta T^{\rm far}\equiv \frac{ \mathcal{R}_{\rm GB} (\theta=\pi)}{ \mathcal{R}_{\rm GB} (\theta=\pi/2)} \ .
\label{deltat_2}
\end{equation}
In the case of~\eqref{eq:GB_perturb}, to $\mathcal{O}(\epsilon)$, 
\begin{align} \label{eq:deltat_result}
\delta T^{\rm close}=\delta T^{\rm far}=\delta T \simeq 1+24\epsilon \ .
\end{align}
%
This quantity informs us, as well, that tides are symmetric and confirms that the HT is indeed larger than the LT (since $\epsilon>0$); unlike~\eqref{deltat} it does not tell us the HT is twice as strong as the LT, but it measures the ``speed" at which HT/LT ratio is departing from unity in the linear approximation.

All of the above descriptions of gravitational tides are in qualitative agreement; but they all also rely on (different) approximations. One may ask how do the tides behave in the fully non-linear regime of GR; namely, how large can the ratios~\eqref{deltat} or~\eqref{deltat_2} be? Additionally, can the LT in a fully non-linear binary in GR become so significant as to change the sign of $\mathcal{R}_{\rm GB}$? 

This last question, besides its academic interest, has significance to the recent programme of BH \textit{spontaneous scalarization} induced by the GB invariant - see~\cite{Doneva:2022ewd} for a review - which provides a dynamical mechanism to challenge the Kerr hypothesis~\cite{Herdeiro:2022yle}, 
leading to non-Kerr BHs - see $e.g.$\til\cite{Silva:2017uqg,Doneva:2017bvd,Witek:2018dmd,Silva:2018qhn,Minamitsuji:2018xde,Doneva:2019vuh,Fernandes:2019rez,Minamitsuji:2019iwp,Cunha:2019dwb,Andreou:2019ikc,Ikeda:2019okp,Ramazanoglu:2017xbl,Doneva:2017duq,Annulli:2019fzq,Kase:2020yhw,Ramazanoglu:2019gbz,Ramazanoglu:2017yun,Ramazanoglu:2018hwk,Minamitsuji:2020hpl,Antoniou:2017acq,Annulli:2021lmn}. It occurs in extended scalar-tensor models, when a real scalar field is non-minimally coupled to gravity through the GB invariant.
According to whether the instability of the vacuum Schwarzschild/Kerr BH is triggered by spacetime regions of positive, or negative, GB invariant, it is dubbed GB$^{+}$ or GB$^{-}$ scalarization, respectively\til\cite{Herdeiro:2021vjo}. GB$^+$ scalarizes both Schwarzschild and Kerr BHs, but it is suppressed for the latter;  GB$^-$ only scalarizes Kerr BHs with dimensionless spins $j>0.5$, thus being dubbed {\it spin-induced}\til\cite{Dima:2020yac,Herdeiro:2020wei,Berti:2020kgk}.

It turns out that GB$^-$ scalarization can also be triggered by other sources rather than the BH spin, {\it e.g.}, self-gravitating magnetic fields\til\cite{Herdeiro:2018wub,Brihaye:2021jop,Annulli:2022ivr}. Since tidal interactions can suppress the GB invariant in LT regions, one may ask whether these could source GB$^-$ scalarization when non-linearities of GR are considered.  To tackle this question we exploit the integrability properties of the Einstein equations~\cite{PhysRevLett.19.1095,PhysRev.167.1175,PhysRev.168.1415} - see also\til\cite{Alekseev:2010mx} and references therein  - which allow the construction of static or stationary binary BHs in equilibrium. Concretely, we shall use the model  in~\cite{Astorino:2021dju,Vigano:2022hrg} that yields a system of one or more \textit{balanced} BHs immersed in an external gravitational field, without any local pathologies, including no conical singularities. 
The external gravitational field can be seen as due to very far away sources, providing an \textit{analytic local} model for the tidal effects of gravitational fields in one or more BHs - in the latter case balanced by tunning the external gravitational field - in \textit{fully non-linear GR}. Additionally, when the motion of the external sources occurs in time-scales large compared to the local dynamics, this metric provides a proxy to a \textit{dynamical} binary.

This paper is organized as follows. In Sec.\til\ref{subsec:Black_holes_in_external_fields}, we  report the line element for multi-BH configurations in an external gravitational multipolar field. Subsecs.\til\ref{subsec:External field with no black hole}-\ref{External field with one black hole}-\ref{subsec:Black_hole_binaries} are dedicated to the analysis of the tides and the GB invariant for a spacetime describing a gravitational field without any BH, and the case of one BH and a BH binary immersed in external fields, respectively. In Sec.\til\ref{sec:Conclusions} we provide final remarks.

\section{Black holes in external fields}\label{subsec:Black_holes_in_external_fields}
\label{sec2}
Consider the following stationary and axi-symmetric metric in Weyl-type  cylindrical coordinates, adapted to these symmetries, such that $\partial_t$ and $\partial_{\phi}$ are Killing vectors,
\begin{align}\label{eq:general_metric}
ds^2=f\left(\rho,z\right)\left(d\rho^2+dz^2\right)+g_{ij}\left(\rho,z\right)dx^i dx^j\,,
\end{align}
where the $i,j$ indices correspond to $t$ or $\phi$ only. This metric form is suitable for both describing static Weyl solutions~\cite{Weyl:1917gp} (see also~\cite{Emparan:2001wk}) and for the application of the inverse-scattering method\til\cite{Belinsky:1971nt} (see also~\cite{Harmark:2004rm}), generating their stationary counterparts. 
With an appropriate choice of the metric functions, Eq.\til\eqref{eq:general_metric} represents an array of $N$ static BHs, free of conical singularities, immersed in an external gravitational multipolar field, generalizing therefore the Israel-Khan solution\til\cite{Israel:1964}. 
The external field is the general relativistic counterpart of the familiar solution in Newtonian gravity (or electrostatics) describing an (axisymmetric) multipolar expansion of the gravitational (or electric) potential in Euclidean 3-space, expressed in standard spherical coordinates. Restricting to the radially growing multipoles, it  reads
\begin{align}
\varphi= \sum_{l=1}^{\infty} b_l \, r^l \, Y^{l0}\left(\theta\right)\, ,
\label{newmul}
\end{align}
where the $b_l$ determine the multipoles' strength.  The gravitational field of~\eqref{newmul}  diverges at $r\rightarrow \infty$. This is because the solution should be truncated at some large $r$ where sources leading to these multipoles should be placed. Likewise, the spacetimes we shall be considering are not asymptotically flat, becoming singular at spatial infinity. This should be interpreted as the theory demaning appropriate source terms to be placed far away to justify the external multipolar gravitational field. With this interpretation, these solutions can be considered as useful proxies for a local analysis, in the vicinity of BHs' horizons, when distant sources perturb their spacetime. 

The metric functions that decribe $N$ static BHs in equilibrium within an external gravitational field, and aligned along the $z$-axis, are the following\til\cite{Vigano:2022hrg}:
\begin{align}
&f^N=16 C_f f_0 \prod_{k=1}^N [\mu_{2k}^{2N+1}\mu_{2k-1}^{2N-1}]\left[\prod_{k=1}^{2N} \frac{1}{\rho^2+\mu_k^2}\right]\nn\\
&\times\prod_{k=1,l=1,3,\dots}^{\stackrel{{{\rm Max}(k+l)=2N}}{2N-1}} \frac{1}{\left(\mu_k-\mu_{k+l}\right)^2}\prod_{k=1,l=2,4,\dots}^{\stackrel{ {{{\rm Max}(k+l)=2N}}}{2N-2}} \frac{1}{\left(\rho^2+\mu_k \mu_{k+l}\right)^2}\nn\\\label{eq:metric_functions}
&\times\exp\left(2 \sum_{k=1}^{2N}\left(-1\right)^{k+1}F\left(\rho,z,\mu_k\right)\right)\nn\\
&g_{ij}^N=\text{diag}\Bigg\{
-\frac{\prod_{k=1}^N \mu_{2k-1}}{\prod_{l=1}^N \mu_{2l}} e^h ,\rho^2 \frac{\prod_{l=1}^N \mu_{2l}}{\prod_{k=1}^N \mu_{2k-1}} e^{-h}
\Bigg\}\,,\nn\\
&h=2 \sum_{n=1}^\infty b_n r^n P_n\,.
\end{align}
In Eqs.~\eqref{eq:metric_functions}, $r\equiv \sqrt{\rho^2+z^2}$; $P_n(z/\sqrt{\rho^2+z^2})$ is the Legendre polynomial of order $n$; $\mu_k\equiv\sqrt{(z-w_k)^2+\rho ^2}-(z-w_k)$, where the $w_i$, called  {\it poles}\til\cite{Vigano:2022hrg}, are related to the mass, and position parameters of each BH as $w_1\equiv z_1-m_1, w_2\equiv z_1+m_1,\dots, w_{2N}\equiv z_N+m_N$; and the functions $C_f$, $f_0$ and $F\left(\rho,z,\mu_k\right)$ are given respectively by
\begin{align}\label{eq:metric_functions_2}
& {C_f=2^{2N^2-4}\prod_{k=1,l=1,3,\dots}^{\stackrel{{\rm Max}(k+l)=2N}{2N-1}} (w_{k}-w_{k+l})^2 \ , } \nn\\
& f_0=\exp\Bigg[\sum_{n,p=1}^\infty \frac{2n p b_n b_p r^{n+p}}{n+p}\left(P_n P_p - P_{n-1}P_{p-1}\right)-h\Bigg]\nn\,,\\
& F(\rho,z,\mu)=2 \sum_{n=1}^\infty b_n\Bigg[\sum_{l=0}^\infty \binom{n}{l} \left(-\frac{\rho^2}{2\mu}\right)^l \left(z+\frac{\mu}{2}\right)^{n-l}\nn\\
&-\sum_{l=1}^n \sum_{k=0}^{(n-l)/2} \frac{\left(-1\right)^{k+l} 2^{-2k-l} n! \mu^{-l}}{k! (k+l)! (n-2k-l)!}\rho^{2(k+l)}z^{n-2k-l}\Bigg]\,,
\end{align}
where $C_f$ is a parameter useful for the conical-singularity regularization\til\cite{Vigano:2022hrg}. Being interested in the value of curvature invariants along the BH horizon, let us stress that, for the $i$-th BH, its horizon is located along the $z$-axis, in the range $w_{2i-1}<z<w_{2i}$. 

Overall, the metric is determined by choosing $N$ (the number of BHs), the external multipole strengths $b_n$ and the constants $z_i,m_i$, $i=1\dots N$,  parameterizing the $i$-BH ``mass", $m_i$, and the position where the BH horizon is centered, $z_i$.  The latter two (sets) are useful to parametrize the problem; however, they do not have a direct physical interpretation. To connect them to physical quantites, one introduces the Komar mass\til\cite{PhysRev.113.934,10.1143/PTP.72.73} and the proper distance between horizons in multi-BH system. The Komar-Tomimatsu mass integral is computed locally on each horizon, but it depends on the normalization of the timelike Killing vector field ($\alpha$). The non-asymptotic flatness of the spacetime implies an ambiguity in this normalization. The Komar mass yields\til\cite{Astorino:2021dju},
\begin{equation}
\label{eq:Komar_mass}
M_i=\alpha \int_{z_i-m_i}^{z_i+m_i} dz [\rho g_{tt}^{-1} \partial_\rho g_{tt} |_{\rho=0}=\alpha m_i\,.
\end{equation}
Thus, the relation between the physical (Komar) mass $M_i$ and the mass parameter $m_i$ depends on an undermined normalization constant which, in principle, can be a function of the $b_n$. We shall see below how tidal information can lead to a determination of this constant to linear order in $b_n$.
Considering two neighbouring BHs, the proper distance between them along the $z$-axis is given by
\begin{equation}
\label{eq:proper_distance}
L= \int_{z_i-m_i}^{z_{i+1}+m_{i+1}} dz \sqrt{g_{tt}} |_{\rho=0}\,.
\end{equation}

Embedding a BH in an external multipolar field provides an infinite set of coefficients (the multipoles $b_n$), that allows for the regularization of the metric. In fact, the metric obtained in this way will be conical-singularity free\til\cite{Astorino:2021dju}, generalizing known regularization conditions  where the BH was fixed at the center of the reference frame\til\cite{PhysRevD.57.3382}. In the next sections we will specialize the above results to the case of zero, one or two BHs in an external gravitational field.

\section{Analysis of the tidal interactions}
\label{sec3}

\subsection{External field with no black hole}
\label{subsec:External field with no black hole}
Let us start by considering the case where only an external field is present, without any BHs. This metric represents the ``background" wherein 
static BHs  can be added to construct the generic spacetime with $N$ BHs in equilibrium. This procedure bears some resemblance to the procedure of  adding BHs inside a Melvin Universe\til\cite{Melvin:1963qx,Ernst:1976mzr,Gibbons:2013yq}, also exploiting the integrability of the Einstein(-Maxwell) equations.

 The line element for the pure multipolar field configuration is given by Eq.\til\eqref{eq:general_metric}, where the generic function $f\left(\rho,z\right)$ is given by $f_0$, $cf.$ Eq.\til\eqref{eq:metric_functions_2}, and
\begin{align}
&g_{ij}^0=\text{diag}\left\{-e^h ,\rho^2 e^{-h}\right\}\,.
\end{align}

Considering up to $b_n=3$, where $b_1,b_2,b_3$ determine the dipolar, quadrupolar and octupolar field strengths', respectively, the GB invariant \textit{along  the $z$-axis} admits a neat compact expression given by,
\begin{align}\label{eq:Rgb_NO_BH}
&\mathcal{R}_{\rm GB}|_{\rho=0}=48 e^{4 z (b_1+z (b_2+b_3 z))} \big[b_1^2+2 b_1 z (2 b_2+3 b_3 z)\nn\\
&+z \left(4 b_2^2 z+3 b_3 \left(4 b_2 z^2+3 b_3 z^3+1\right)\right)+b_2\big]^2\,.
\end{align}
Clearly,  the GB invariant is always non-negative along the $z$-axis. Moreover, we have checked, performing a numerical scan, that the GB invariant remains positive semi-definite outside the $z$-axis as well. Notably, this result contrasts with the case of a pure magnetic (Melvin) universe, that to some extent can be considered an electrovacuum counterpart of the multipolar gravitational field described here (but with some relevant differences, namely being non-singular at infinity) where the $\mathcal{R}_{\rm GB}$ can take negative values  sufficiently far away from the center\til\cite{Brihaye:2021jop,Annulli:2022ivr}.

Eq.\til\eqref{eq:Rgb_NO_BH} shows a new (curious) feature of these spacetimes; given a pair of non-zero $\{b_2,b_3\}$ coefficients, there will be always a non-zero value of $z$ such that the GB invariant vanishes. This peculiar property will be inherited by the cases below, including  BHs.  Although the significance of this (zero measure) region with vanishing GB is not clear, it is worth pointing out that it is not exclusive of  a fully relativistic description of these spacetimes. In fact,  the following illustration shows that these zero-GB points have a counterpart within a Newtonian approach to tidal fields. 

Let us introduce a Newtonian gravitational potential $\varphi$ decomposed into spherical harmonics, within a $3D$ flat metric in spherical coordinates. As mentioned above, this external field would represent the effect of far away sources, in the same spirit of the fully general relativistic solutions we are describing. To sharpen the parallelism with the relativistic case, let us also consider that the external sources generating the tidal field are axisymmetric, thus $m=0$ in the spherical harmonics decomposition, precisely given by~\eqref{newmul}.
Defining the Newtonian tidal tensor as $E_{ij}=\nabla_i \nabla_j \varphi$, and considering up to $l=3$, the tidal invariant $E_{ij}E^{ij}$ computed at $\theta=0$, becomes
\begin{align}
E_{ij}E^{ij}\propto 63 b_3^2\left( r+\frac{1}{3}\sqrt{\frac{5}{7}}\frac{b_2}{b_3} \right)^2\,,
\label{newtidal}
\end{align}
which admits a double root whenever $r=-\frac{1}{3}\sqrt{\frac{5}{7}}\frac{b_2}{b_3}$. 

Since the GB invariant  is defined through contractions of the Riemann tensor, which in turn represents a relativistic version of the Newtonian tidal tensor, \eqref{newtidal} can be considered a Newtonian counterpart of~\eqref{eq:Rgb_NO_BH}, where one observes the same qualitative feature that there are particular points along the axis where the tidal invariant vanishes. Thus, these points are non specific to GR, as they also appear in Newton's theory. There are, however, important differences. Of relevance, the $b_1$ term in the Newtonian case does not enter the tidal tensor, as this corresponds to a constant acceleration field, and thus  with vanishing tidal forces. It would be interesting to provide a deeper physical interpretation of these special points.

\subsection{External field with one black hole}
\label{External field with one black hole}
Next we consider a single BH immersed in an external multipolar field, given by\til\eqref{eq:general_metric}, \eqref{eq:metric_functions} and \eqref{eq:metric_functions_2}  with $N=1$. 
We divide our analysis into two cases. First we consider the multipolar expansion of the external field only up to quadrupolar order; then we also add the octupole. In either case, these multipoles can be interpreted as leading order terms of a putative external field, due to a far away source, possible a second BH, but considered fully \textit{non-linearly}. We can then analyse how the (fiducial) BH in our solutions responds. 

The external field expanded up to the quadrupolar order surrounding a Schwarzschild BH translates into considering $b_n=0$, $n>2$ and (possibly) only $b_1,b_2\neq 0$,  in Eq.\til\eqref{eq:metric_functions}. A BH in such external gravitational field will, in general, accelerate. Thus, the generic static solution has conical singularites, reflecting the extra force necessary to fix the BH.  These singularities can be avoided by tunning the dipole and quadrupole, via the conditions~\cite{Vigano:2022hrg}
\begin{align}\label{eq:regularization_1_BH}
&C_f=\frac{\left(w_1-w_2\right)^2}{4}\,,\quad\sum_{n=1}^\infty b_n \left(w_1^n-w_2^n\right)=0\,.
\end{align}
For the case with only  $b_1,b_2\neq 0$, Eq.\til\eqref{eq:regularization_1_BH} yields the simple relation: 
\begin{equation}
b_1=-2z_1 b_2 \ ,
\label{b1b2}
\end{equation}
where $z_1$ determines the position of the BH on the $z$-axis. A special case is $z_1=0$ and $b_1=0$, which amounts to a \textit{purely quadrupolar} field. This possesses a nice interpretation. A BH placed at the symmetric point $z_1=0$  of a purely quadrupolar field does not accelerate. Hence, no dipole is necessary to counteract the acceleration. Additionally, $b_2$ is a free parameter. Thus this case with  $z_1=0$ and $b_1=0$ can be thought as an \textit{extension into the non-linear regime} of the perturbative geometry considered in the Introduction - see Eq.~\eqref{eq:metric_func_schw_tidal}. This viewpoint will be explored  and corroborated below, in particular to identify the physical mass of the non-perturbative geometry.

For $b_1,b_2\neq 0$ and $z_1$ generic, the GB invariant for this configuration, \textit{on the BH horizon}, can again  be expressed in a neat compact form:
\begin{align}\label{eq:Rgb_1_BH}
&\mathcal{R}_{\rm GB}|_{hor}=\frac{3}{4 m_1^4}\Big[1+ 2 b_2 G_1\Big]^2 \exp(4 b_2 G_2)\,,\nn\\
&G_1=  m_1^2-4 m_1^2 b_2 (z-z_1)^2+(z-z_1)^2 \left(4 b_2 (z-z_1)^2-5\right)\nn\\
&G_2=z (z-2 z_1)+(m_1-z+z_1) \left| m_1+z-z_1\right| \nn\\
&+(m_1+z-z_1) \left| m_1-z+z_1\right|\,,
\end{align}
where $z$ spans the interval $]z_1-m_1,z_1+m_1[$ and $m_1$ is the parameter related to the BH's Komar mass, $cf.$~\eqref{eq:Komar_mass}. Only the quadrupole coefficient $b_2$ enters this expression, as the dipole has been replaced by the regularization condition~\eqref{b1b2}. In the limit of vanishing external field, Eq.\til\eqref{eq:Rgb_1_BH} correctly reduces to the single Schwarzschild case $\mathcal{R}_{\rm GB}|_{hor}={3}/(4 m_1^4)$, where $m_1$ can be identified with the physical mass. 

The compact result obtained in Eq.\til\eqref{eq:Rgb_1_BH} clearly shows that the GB invariant is positive semi-definite, meaning that the necessary condition for GB$^-$ scalarization is never met, at least locally in the vicinity of the horizon. This is where strong-field effects are mostly expected to occur. From another viewpoint,  the inclusion of the external gravitational fields appears to  enhance GB$^+$ scalarization, in the HT regions.
\begin{figure}
\centering
\includegraphics[width=8.6cm,keepaspectratio]{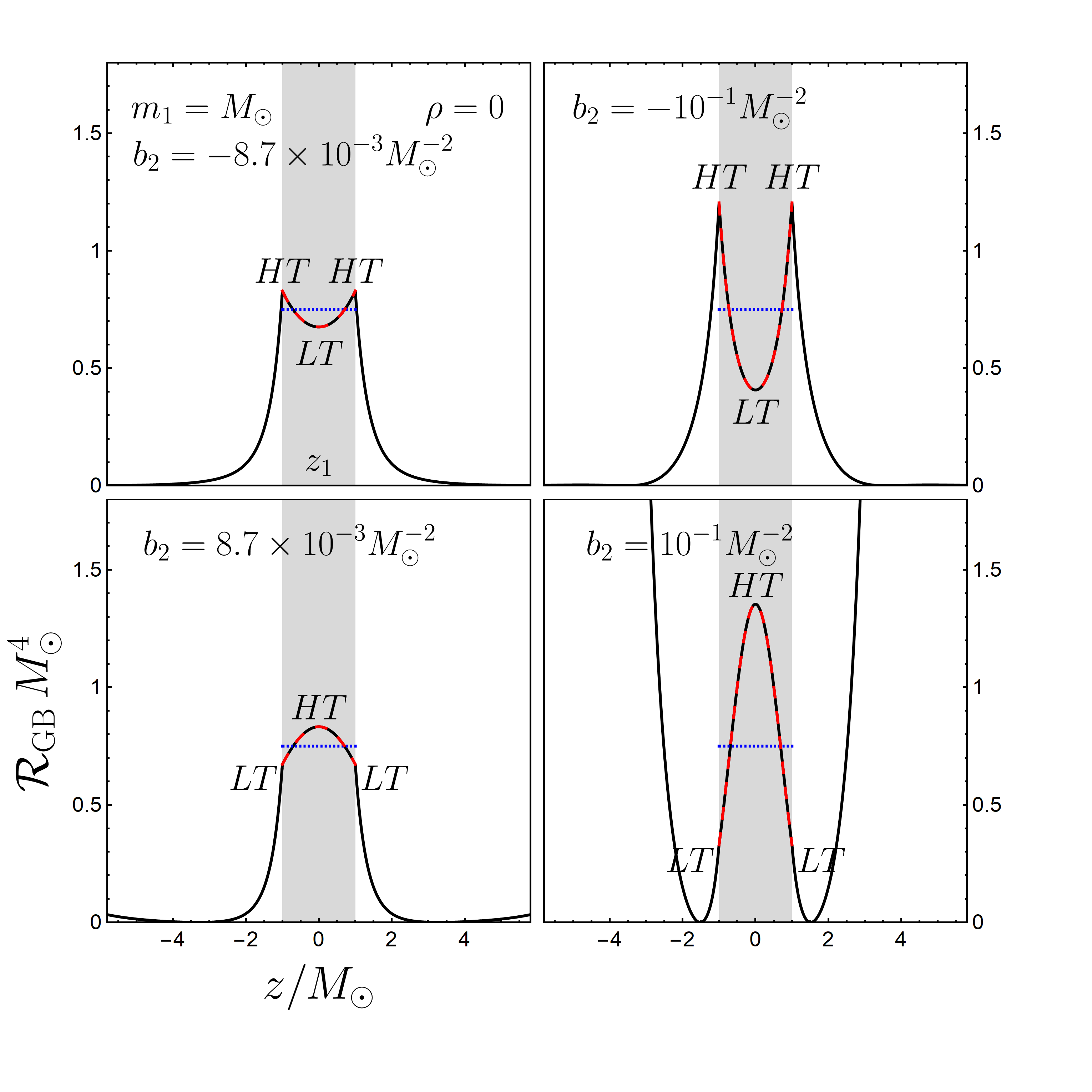} 
\caption{GB invariant along the $z$-axis ($\rho=0$), for a BH in a purely quadrupolar external field ($z_1=0$, hence $b_1=0$, $cf.$ Eq.\til\eqref{b1b2}), with mass $m_1=1 M_\odot$ and two negative -  {\bf upper panels} - or positive -  {\bf lower panels} -   $b_2$ values. The {\bf red, dashed} lines follow the analytic formula in Eq.\til\eqref{eq:Rgb_1_BH}, valid only on the horizon. For $b_2<0$, the horizon region shows, qualitatively, the same HT/LT regions as in Fig.~\ref{fig:plotRGB_perturbation} for a first order perturbed BH; but now this is an exact result in GR. The {\bf blue, dotted} line represents the GB invariant for the unperturbed BH ($b_2=0$) with $m_1=1M_{\odot}$.
}
\label{fig:RGB_1BH_out_horizon}
\end{figure}
In Fig.\til\ref{fig:RGB_1BH_out_horizon} we show how the GB invariant changes along the $z$-axis, including both the horizon and the axis outside the horizon, in the same spirit as Fig.~\ref{fig:plotRGB_perturbation}, confirming that $\mathcal{R}_{\rm GB}$ is positive semi-definite, along the $z$-axis. Inspection of  Fig.~\ref{fig:plotRGB_perturbation} reveals, moreover, that the sign of $b_2$ determines where the HT/LT regions are. Whereas $b_2<0$ (corresponding to a positive quadrupole~\cite{Vigano:2022hrg}) gives a qualitatively similar picture to the perturbative case, with the HT regions in the vicinity of the poles (and the symmetry axis), for $b_2>0$ these become the LT regions, with the HT region along the equator. A more quantitative description is provided next, using the HT/LT ratios introduced above.

\subsubsection{The HT/LT differential ratio}
\label{The mass of the (un)perturbed configuration}

Let us consider first the HT/LT differential ratio~\eqref{deltat}. To compute it, we need to  clarify {\it which} is the unperturbed configuration associated with a BH of mass $m_1$, immersed in a tidal field with given quadrupole coefficient $b_2$ at a position $z_1$ along the $z$-axis.

Setting $b_2=0$ in Eq.\til\eqref{eq:Rgb_1_BH} leads to $\mathcal{R}|_{hor}=3/(4m_1^4)$, the GB invariant of an isolated Schwarzschild BH of physical mass  $M=m_1$. For the non-linearly perturbed BH within the tidal field, however, in general, $m_1= m_1(M, b_2, z_1) \neq M$, with $m_1(M,0,z_1)=M$. This is another view on the ambiguity in choosing $\alpha$ in~\eqref{eq:Komar_mass}.

Interestingly, $M(m_1, b_2, z_1)$ can be determined to linear order in $b_2$ and for $z_1=0$ by comparing the tidal tensors of the fully non-linear solution with the perturbative analysis in the Introduction. For this case, we can expand $m_1$ in powers of the dimensionless quantity $M^2b_2$ in the vicinity of $b_2=0$,
\begin{align}\label{eq:mass_expansion}
m_1 =  M \left(1+ a_1 M^2 b_2 + a_2 (M^2 b_2)^2 \dots\right)\,,
\end{align}
with unknown dimensionless coefficients $a_1,a_2$.
From Eq.~\eqref{eq:Rgb_1_BH} we find that the GB invariant at the poles ($ z=\pm m_1$) and equator ($z=0$), to linear order in $b_2$ are, respectively, 
\begin{align}
\label{eq:rgb_nonlinear}
&\mathcal{R}|_{hor}=\frac{3}{4m_1^4}-\frac{9b_2}{m_1^2}\stackrel{\eqref{eq:mass_expansion}}{=}\frac{3}{4M^4}-\frac{3b_2(a_1+3)}{M^2}\,,  \nn\\
&\mathcal{R}|_{hor}=\frac{3}{4m_1^4}+\frac{9b_2}{m_1^2}\stackrel{\eqref{eq:mass_expansion}}{=}\frac{3}{4M^4}-\frac{3b_2(a_1-3)}{M^2} \,,
\end{align}
where the last equalities implement the relation between the mass parameter $m_1$ and the physical mass $M$ to linear order in $b_2$, $cf.$~\eqref{eq:mass_expansion}.
If the non-linear solution of a BH within the quadrupolar field provides a good local description of tidal effects, then \eqref{eq:rgb_nonlinear} must match the linear perturbation theory treatment (Eq.\til\eqref{eq:GB_perturb}), which we rewrite here as 
\begin{align}
\label{eq:rgb_linear}
&\mathcal{R}|_{hor}=\frac{3}{4M^4}+\frac{12 \epsilon}{M^4}\,, \quad \ \text{at \ the \ poles }\, \nn \\
&\mathcal{R}|_{hor}=\frac{3}{4M^4}-\frac{6 \epsilon}{M^4}\,, \quad \text{ at \ the  \ equator.}
\end{align}
Equations\til\eqref{eq:rgb_nonlinear} and \eqref{eq:rgb_linear} can indeed be matched taking
\begin{equation}
a_1=1 \ ,  \quad \epsilon=-b_2 M^2 \ .
\label{a1b2}
\end{equation}
The fact that $b_2$ takes the opposite values of a physical quadrupole moment is in agreement with Ref.\til\cite{Vigano:2022hrg}. Thus, as a result of matching tidal effects in the fully non-linear GR solution of a BH immersed in a quadrupolar field and a perturbed Schwarzschild BH, we get, for the former, a simple relation, valid for small $b_2$, between the physical mass of the BH at $z_1=0$ and its mass parameter $m_1$:
\begin{align}
m_1 \simeq  M \left(1+  M^2 b_2 \right)\, .
\label{mM}
\end{align}
Substituting Eq.\til\eqref{a1b2} in Eq.\til\eqref{eq:rgb_nonlinear}, we compute the the HT/LT differential ratio\til\eqref{deltat} of the BH immersed in the quadrupolar field to  first order in $b_2$ yielding $\Delta T=-2$, thus matching the linear, non-linear and Newtonian treatments. A computation of $\Delta T$ beyond linear order in $b_2$ requires obtaining the physical mass to that order, which remains an interesting challenge.

\subsubsection{The HT/LT ratio}
\label{The tidal differential}

\begin{figure}[h!]
\centering
\includegraphics[width=6.8cm,keepaspectratio]{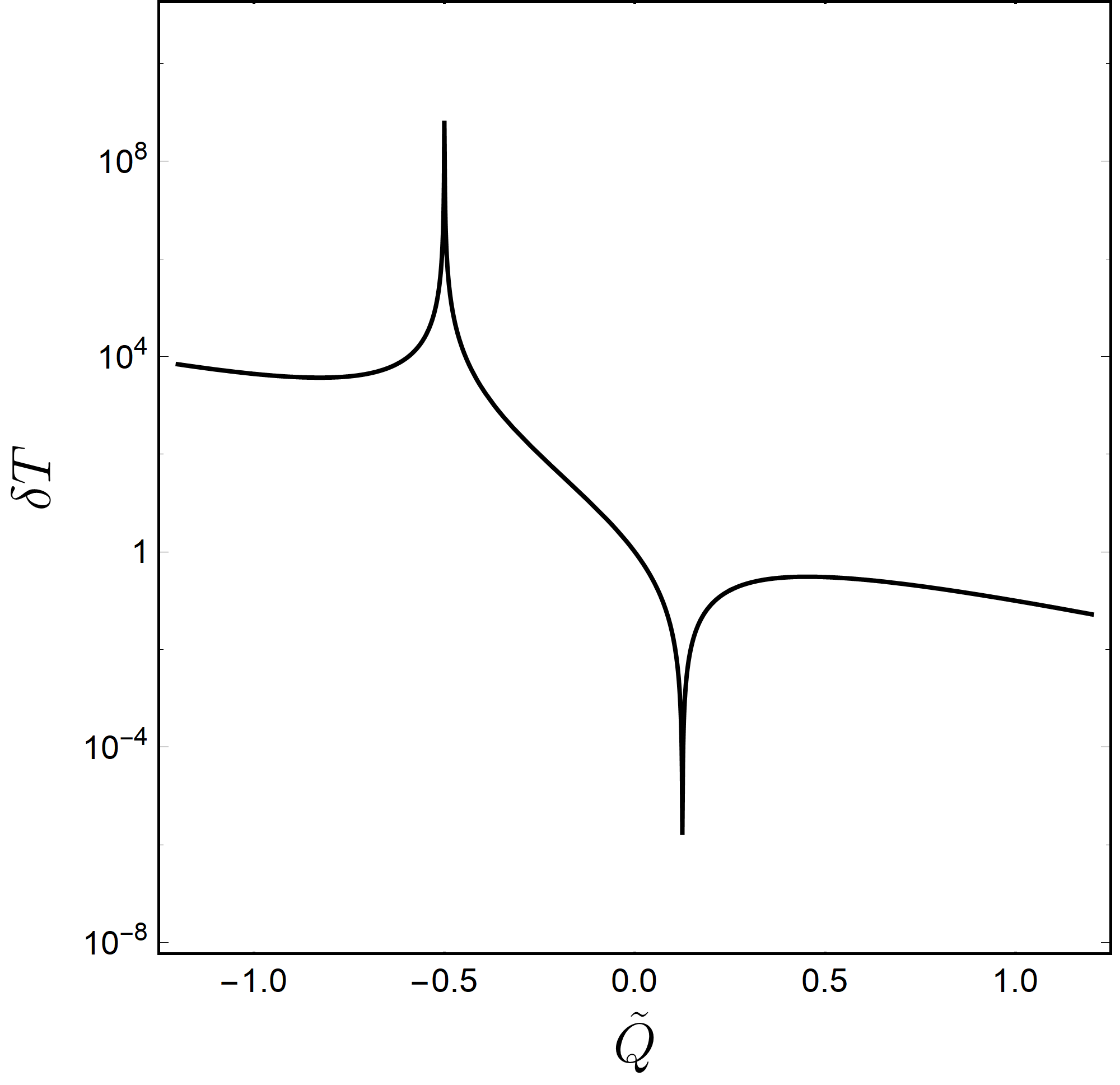} 
\caption{HT/LT ratio - Eq.\til\eqref{deltat_2} - as a function of the dimensionless quadrupole parameter $\tilde{Q}$ for a BH immersed in a dipole-quadrupole field.
}
\label{fig:deltaT_vs_tildeQ}
\end{figure}

Let us now consider  the HT/LT ratio given by Eq.\til\eqref{deltat_2} for the non-linear solution of a BH immersed in a multipolar field. This quantity does not depend on $z_1$; additionally it does not require a comparision with an unperturbed configuration; thus it can be computed \textit{exactly} for the non-linear solution of a BH immersed in the dipole+quadrupolar field. Using~\eqref{eq:Rgb_1_BH} in Eq.\til\eqref{deltat_2} we get,
\begin{align}
\delta T=\frac{(1-8 \tilde{Q})^2}{(2 \tilde{Q}+1)^2} e^{-4 \tilde{Q}} \,, \qquad \tilde{Q}\equiv  b_2 m_1^{2} \ .
\label{tratio1}
\end{align}
 In the small $\tilde{Q}$ limit, we get 
$$\delta T = 1- 24\tilde{Q} + \mathcal{O}\left(\tilde{Q}^{2}\right) \ , $$ that coincides with the linear result - Eq.\til\eqref{eq:deltat_result} -  using eqs.~\eqref{a1b2} and~\eqref{mM}. To illustrate the behaviour of $\delta T$, in Fig.\til\ref{fig:deltaT_vs_tildeQ} we plot it against $\tilde{Q}$.

Fig.\til\ref{fig:deltaT_vs_tildeQ} highlights three different regime of the tidal ratio~\eqref{tratio1} which turn out to be related in a clear way to the \textit{2D geometry} of the spatial sections of the BH horizon. To establish this, we construct an isometric embedding of the spatial section of the BH horizon in $\mathbb{E}^3$, using standard techniques - see $e.g.$~\cite{Smarr:1973zz,Gibbons:2009qe,Delgado:2018khf} - for several values of $\tilde{Q}$, as shown in Fig.\til\ref{fig:embeds}. One observes that:
\begin{figure}
\centering
\includegraphics[width=8.5cm,keepaspectratio]{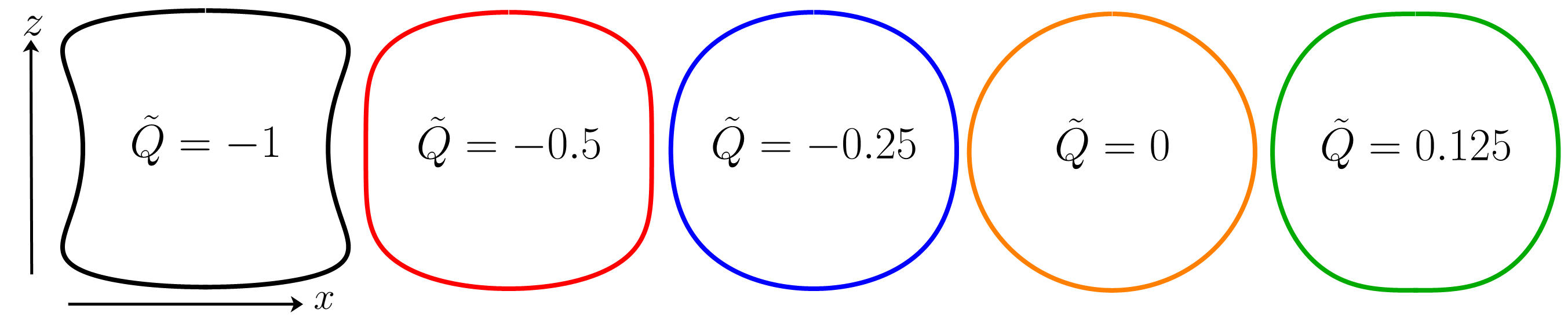} 
\caption{$x$-$z$ section of the  $\mathbb{E}^3$ horizon embeddings, for several values of the dimensionless quadrupole parameter $\tilde{Q}$. The specific coordinates values of each horizon embedding are scaled to show them all in the same range.
}
\label{fig:embeds}
\end{figure}

\begin{enumerate}
\item in the regime  $-1/2<\tilde{Q}<1/8$ the Gaussian curvature of the spatial sections of the horizon is always positive, and the BH horizon is always embeddable in Euclidean 3-space. In the lower end this interval, $\tilde{Q}\simeq -1/2$, the  curvature tends to zero on the equator - Fig.\til\ref{fig:embeds}, $\tilde{Q}=-0.5$ - and $\delta T >1$. In the upper end,  the Gaussian curvature tends to zero at the poles - Fig.\til\ref{fig:embeds}, $\tilde{Q}=0.125$ - and $\delta T<1$.
\item In the regime $\tilde{Q}<-1/2$, the Gaussian curvature of the spatial sections of the horizon becomes negative in the vicinity of the equator, leading to a hyperbolic shape in that neighborhood. An embedding in $\mathbb{E}^3$ can, nonetheless, be performed - Fig.\til\ref{fig:embeds}, $\tilde{Q}=-1$. $\delta T>1$ always.
\item Finally, in the regime  $\tilde{Q}>1/8$ the BH horizon is not embeddable anymore, as its Gaussian curvature becomes negative at the fixed points of the rotation symmetry, $i.e.$, the poles. This is precisely the same phenomenon that occurs for the Kerr horizon for its dimensionless spin larger than $\sqrt{3}/2$~\cite{Smarr:1973zz}.  $\delta T<1$ always.
\end{enumerate}

Let us close this discussion stressing that the 2-dimensional (Gaussian) curvature of the spatial sections of the horizon does become negative in some regions, but the 4-dimensional GB curvature, on the horizon, never does so.

\subsubsection{Area and entropy}
\label{Area and entropy}

Given the relation\til\eqref{mM} between $m_1$ and the physical mass found for small $b_2$, let us consider the area and therefore the entropy of these BHs.

The entropy of a BH in GR is related to the BH area by the Bekenstein-Hawking formula\til\cite{Bardeen:1973gs}. The entropy of the linearly-perturbed BH described by Eqs.\til\eqref{eq:metric_func_schw_tidal} is therefore simply
\begin{align}
S=A/4=4 \pi M^2=S_0\,.
\end{align}
When the BH is instead immersed in the non-linear tidal quadrupolar field, the entropy can be computed through
\begin{align}
\label{eq:BH_entropy_nonlin}
S= \lim_{\rho\rightarrow 0} \int_0^{2\pi}  \int_{-m_1}^{m_1} d\phi dz \sqrt{g_{zz} g_{\phi\phi}}=4 \pi m_1^2 e^{-2 \tilde{Q}^2}\,,
\end{align}
where the BH's location is assumed to be $z_1=0$. In the small $b_2$ limit, and substituting the physical mass through Eq.\til\eqref{mM}, we get that the BH entropy is modified only at second order in the quadrupolar field,
\begin{align}\label{eq:BH_entropy_nonlin_2}
S=S_0 \left(1- (5-2a_2) M^4 b_2^2\right)\,.
\end{align}
Eq.\til\eqref{eq:BH_entropy_nonlin_2} is quadratic in $b_2$, thus containing the unknown $a_2$; depending on its value it can increase or decrease. Thus, knowledge of $a_2$ is needed  to establish the sign of the entropy variation when the BH is immersed in the non-linear tidal field. But the result does confirm that the isolated BH entropy is unaffected at linear $b_2$ order.

\subsubsection{The GB invariant away from the horizon}
\label{The tidal differential}

\begin{figure}
\centering
\includegraphics[width=8cm,keepaspectratio]{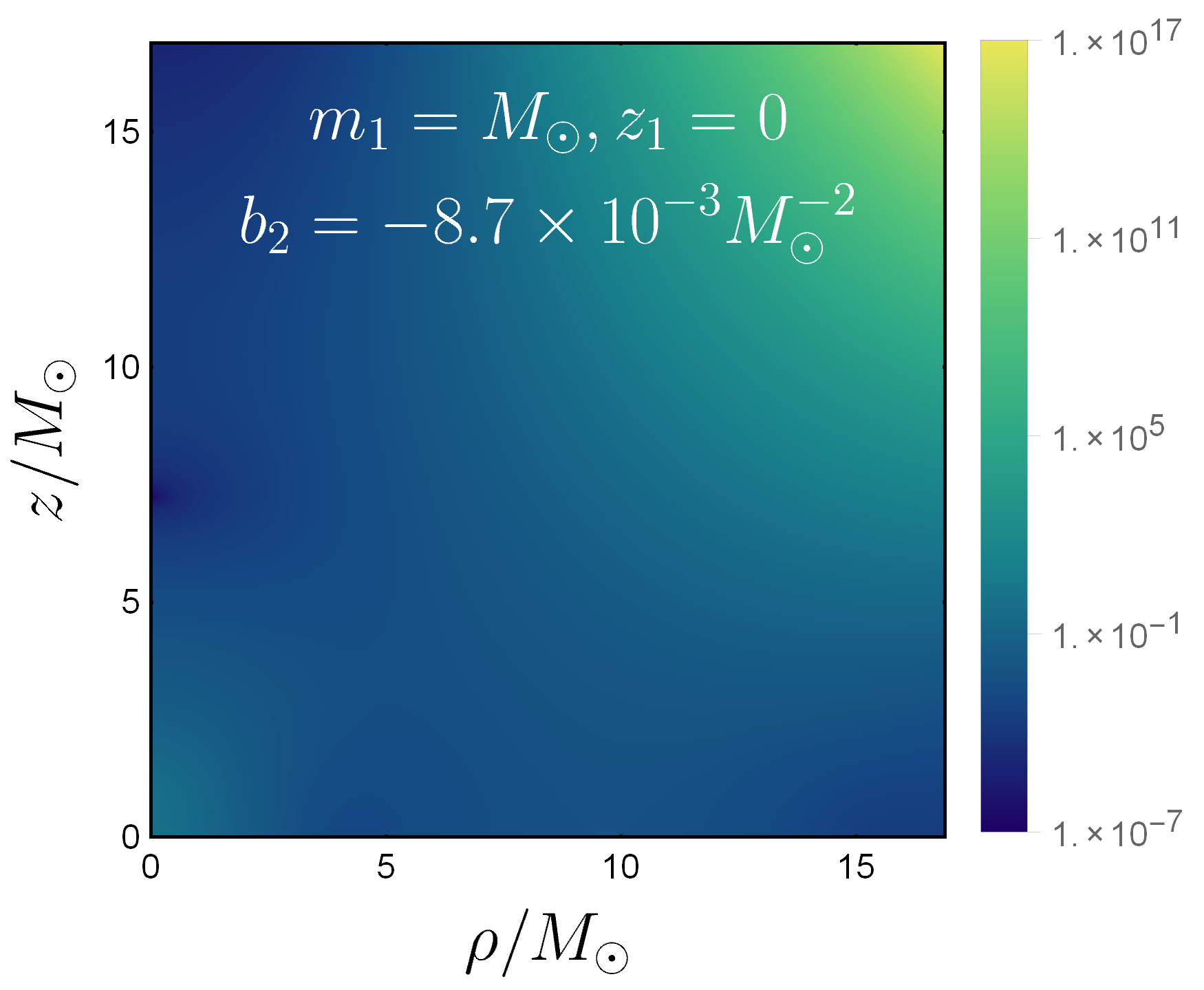} 
\caption{GB invariant for one of the BHs in Fig.\til\ref{fig:RGB_1BH_out_horizon}, on the $\rho-z$ plane. This plot clearly shows how $\mathcal{R}_{\rm GB}$ depends on the specific chosen direction, but it never becomes negative.}
\label{fig:RGB_1BH_out_horizon_far}
\end{figure}

Fig.\til\ref{fig:RGB_1BH_out_horizon} (top panels, which are for $b_2<0$) may suggest that the GB invariant tends asymptotically to zero, as in any asymptotically flat spacetime. But the bottom panels, which are for $b_2>0$, exhibit a growth towards larger $|z|$. In fact, even for $b_2<0$, the GB invariant, as well some of the components of the metric, may either tend to zero or to infinity in the considered geometry, depending on the specific direction in the $\rho-z$ plane. To observe this, in Fig.\til\ref{fig:RGB_1BH_out_horizon_far} we show the GB dependence on the $\rho-z$ coordinates for  one of the examples in Fig.\til\ref{fig:RGB_1BH_out_horizon} with $b_2<0$.

\subsubsection{Adding the octupole}

\begin{figure}
\centering
\includegraphics[width=8.2cm,keepaspectratio]{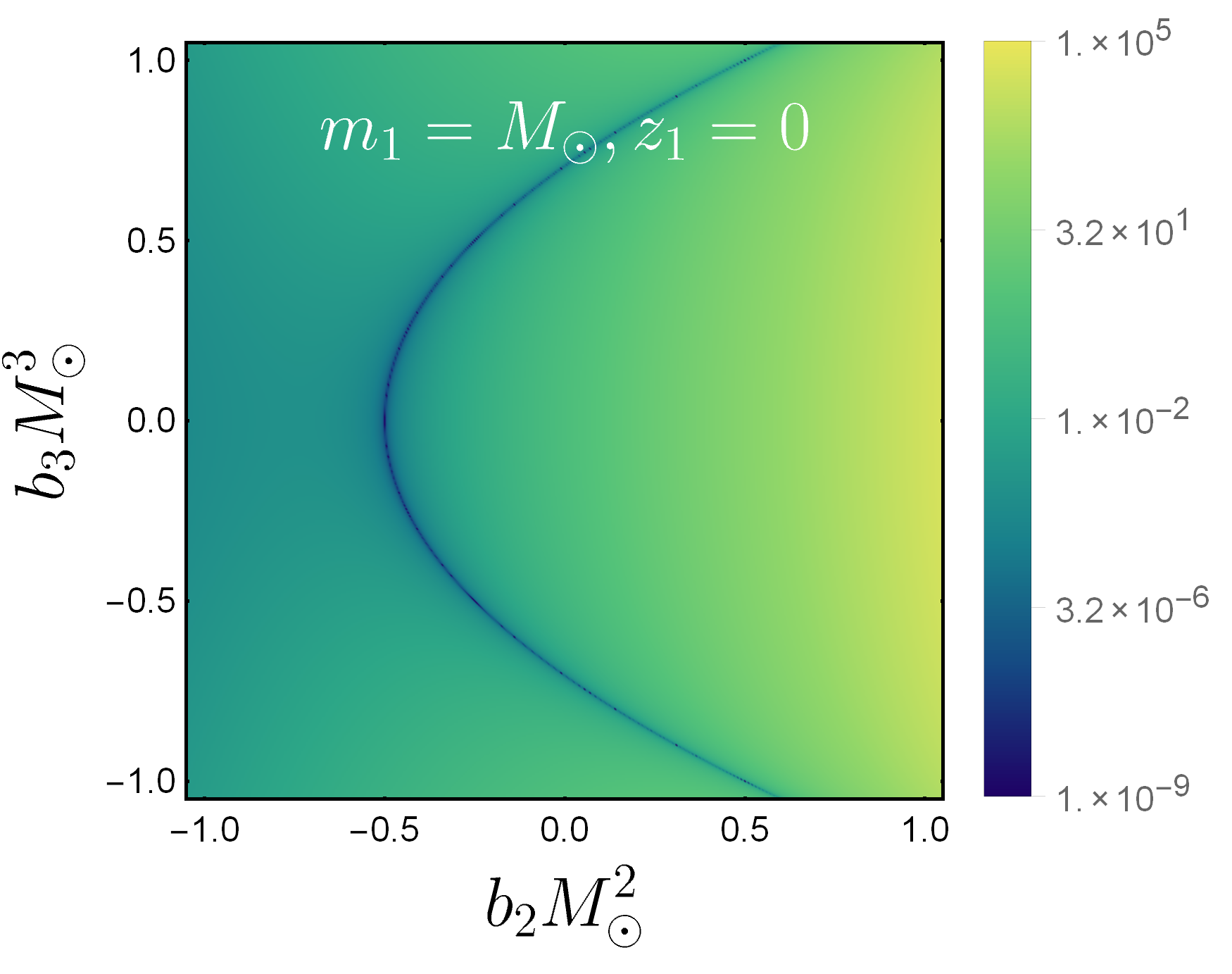} 
\caption{$\mathcal{R}_{\rm GB}$ on the horizon of a BH with the external field up to the octupolar term, in units of $M_{\odot}$. The GB invariant in this plot has been evaluated on the horizon, at a choosen point (the equator, at $z=0$). 
The {\bf dark blue} line corresponds to the zeros of the GB invariant in the $b_2$-$b_3$ space, akin to the roots found in the case  with no BH (Subsec.\til\ref{subsec:External field with no black hole}).}
\label{fig:RGB_1_BH}
\end{figure}

To check the generality of the above discussion concerning the sign of the GB invariant with two multipoles, we next consider an additional degree of freedom, the octupole, taking up to $l=3$ in the multipolar external gravitational field. Thus we take $b_n=0$ for $n>3$ and (possibly) only $b_1,b_2,b_3\neq 0$. The updated regularization conditions\til\eqref{eq:regularization_1_BH}, now determines the dipolar field in terms of the quadrupole and the octupole, as 
$$b_1=-2z_1 b_2 - \left(m_1^2+3z_1^2\right)b_3 \ .
$$ 
Unfortunately, in this more general case no compact, closed form expression for $\mathcal{R}_{\rm GB}|_{hor}$ could be obtained, as in Eq.\til\eqref{eq:Rgb_1_BH}.  We verified numerically, however, that the GB invariant is always positive or null on the horizon of the BH, {\it i.e.}, we could not find a set of parameters $\{m_1,z _1, b_2,b_3\}$ for which $\mathcal{R}_{{\rm GB}}$ takes negative values. As an illustration, Fig.\til\ref{fig:RGB_1_BH} shows a scanning in the $b_2$-$b_3$ space for a particular choice of $m_1,z_1$, and focusing on the horizon point with $z=0$ (the equator). A distinguishable feature is the line of values for which the GB vanishes at this given horizon point ($z=0$), similarly to the discussion before in the case without BHs.
%

%

\subsection{Black hole binaries}
\label{subsec:Black_hole_binaries}
Finally, we turn to the case of a static BH binary immersed in an external multipolar field. The corresponding metric is\til\eqref{eq:general_metric}, \eqref{eq:metric_functions} and \eqref{eq:metric_functions_2}  with $N=2$.
Here, we shall consider a dipolar-quadrupolar external gravitational field, thus with $b_n=0$, $n>2$ and (possibly) only $b_1,b_2\neq 0$.

The regularization conditions in the binary case translates into a set of algebraic conditions relating the masses and positions parameters of the BHs with the multipolar coefficients of the external field. Then, the freedom of choosing $m_i$ and $z_i$ will automatically fix the external field coefficients. More quantitatively, we get
\begin{align}\label{eq:regularization_2_BH}
&\frac{C_f}{{256}}=m_1^2 m_2^2 (m_1+m_2-z_1+z_2)^2(m_1+m_2+z_1-z_2)^2\\
&b_1=\frac{(m_1 z_1+m_2 z_2) \log \frac{(m_1-m_2+z_1-z_2) (m_1-m_2-z_1+z_2)}{(m_1+m_2+z_1-z_2) (m_1+m_2-z_1+z_2)}}{4 m_1 m_2 (-z_1+z_2)}\,,
\label{b12bh}\\
&b_2=\frac{(m_1+m_2) \log \frac{(m_1-m_2+z_1-z_2) (m_1-m_2-z_1+z_2)}{(m_1+m_2+z_1-z_2) (m_1+m_2-z_1+z_2)}}{8 m_1 m_2 (z_1-z_2)}\,.
\label{b22bh}
\end{align}
Making the ratio of~\eqref{b12bh} and~\eqref{b22bh} and taking $m_2=0$ one recovers the single BH condition, Eq.~\eqref{b1b2}.
Physically, the two BHs are both attracted to one another and accelerated by the external fields, so that a generic static solution has conical singularities, emulating the extra force necessary to enforce equilibrium. Tunning the multipolar fields - as in the regularization conditions above - provides an equilibrium without extra forces.  The most intuitive case, in which only the quadrupole is necessary, is the $\mathbb{Z}_2$ symmetric configuration with $m_1=m_2$ and $z_1=-z_2$, leading to no dipole $b_1=0$ and
\begin{equation}
b_2=\frac{ \log \frac{z_1^2}{z_1^2 -m_1^2}}{8 m_1 z_1}\,.
\end{equation}
Thus, unlike the single BH case, for which the $\mathbb{Z}_2$ symmetric configuration had no dipole and an arbitrary quadrupole, in the two BH case,  the $\mathbb{Z}_2$ symmetric configuration has no dipole but a quadrupole that is determined by the mass a distance parameters.

Once a regular equilibrium is guaranteed, the scenario is now that each of the two BHs is subjected both to the external multipolar tidal field (of far away sources) and to the close by tidal field (of the other BH). In particular, this yields the opportunity to analyse the fully non-linear effect of a close by perturber. Thus, we treat one of the BHs as the fiducial one and the other as the (non-linear) perturber. Obviously, the assignment of each of the roles to each of the BHs is arbitrary.

\begin{figure}
\centering
\includegraphics[width=7cm,keepaspectratio]{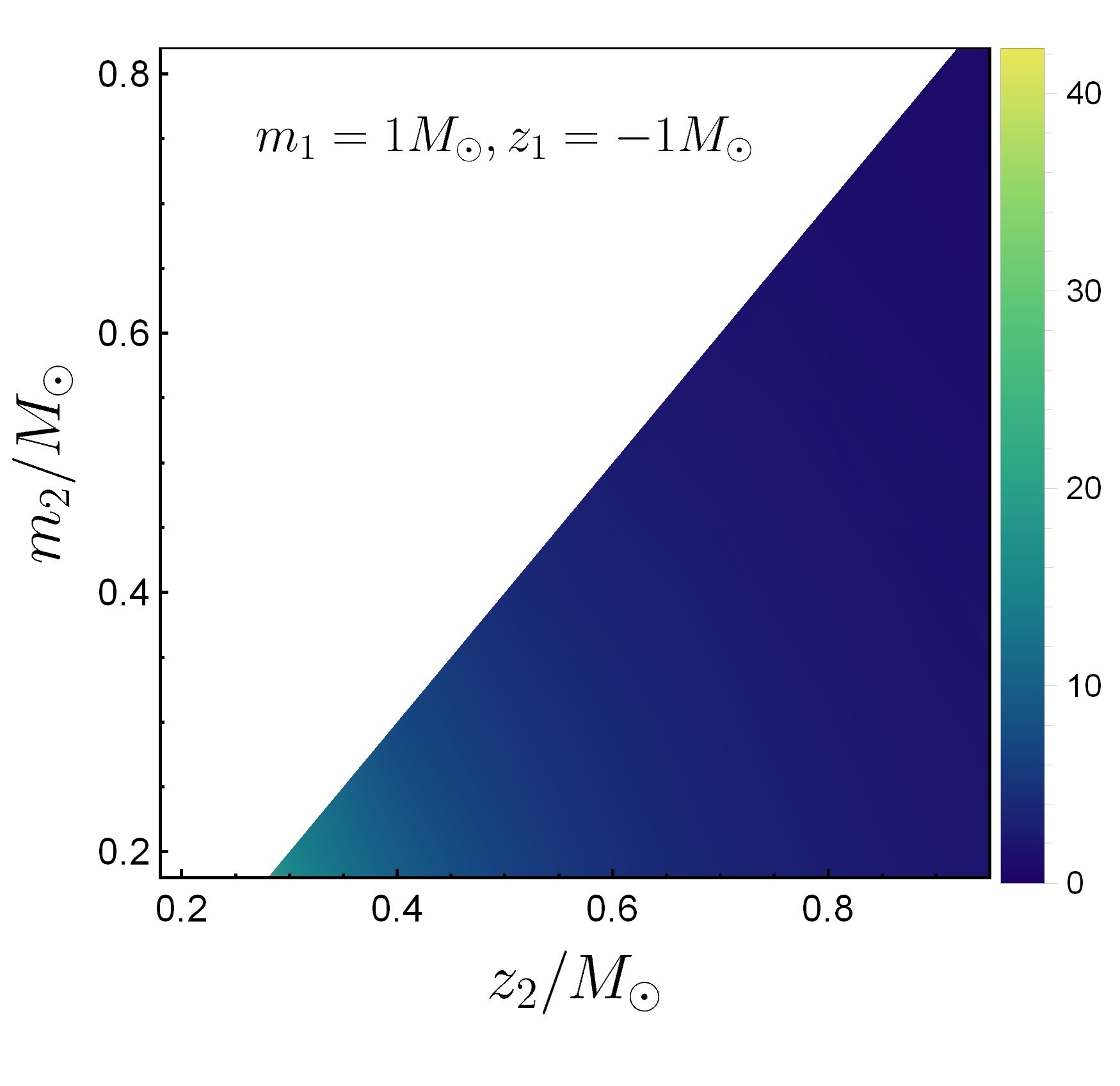} \includegraphics[width=7cm,keepaspectratio]{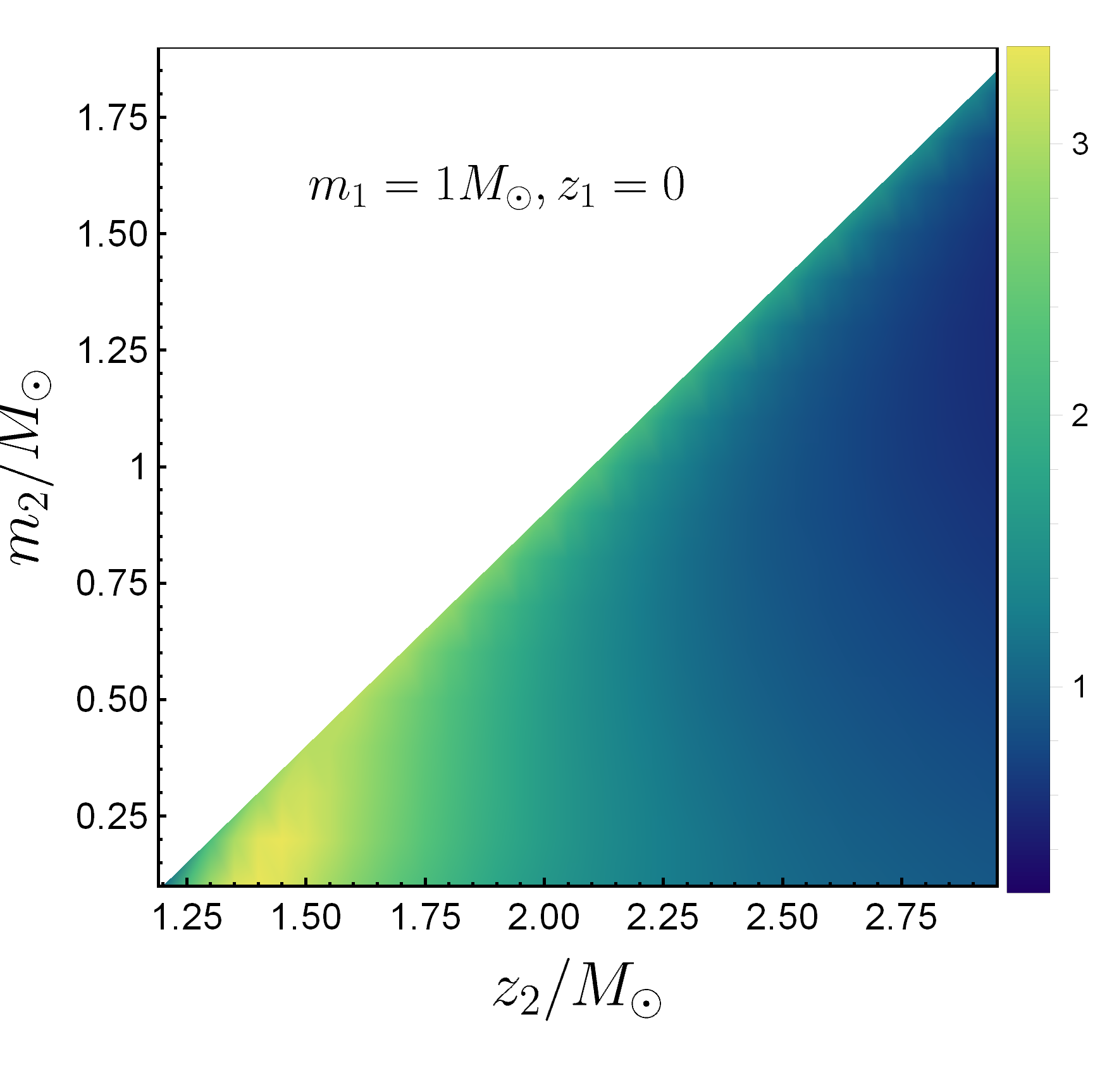} \includegraphics[width=7cm,keepaspectratio]{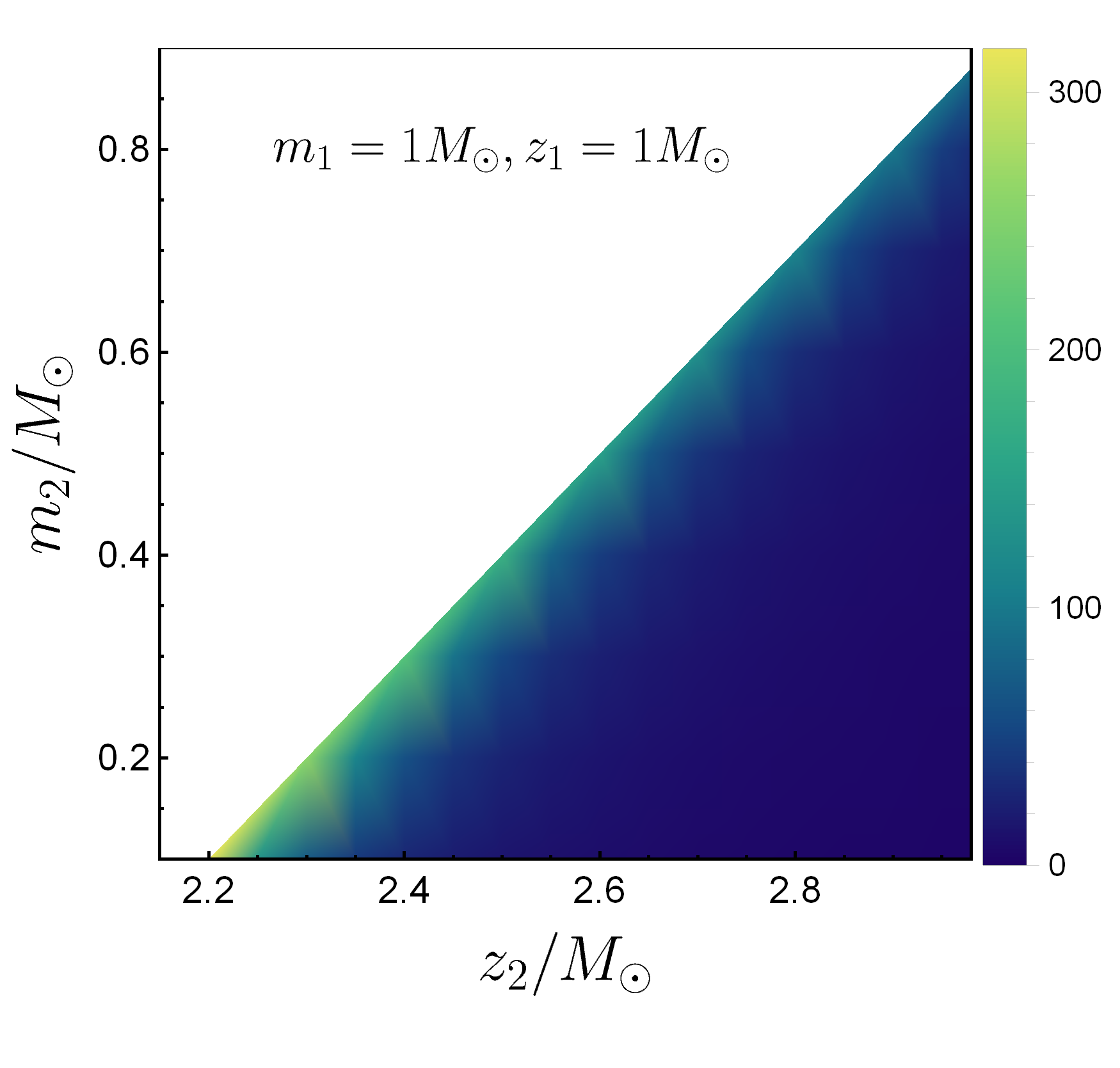} 
\caption{The GB invariant on the horizon of the fiducial BH in the binary, with external dipolar and quadrupolar field, in units of $M_{\odot}$. In each planel, $\mathcal{R}_{\rm GB}$ is evaluated at $z=z_1+m_1$ and $\rho=0$, fixing the first (fiducial) BH mass/position parameters ($m_1,z_1$), as shown on the panel. The white portion in each panel represents configurations in which the second BH (perturber) horizon would overlap with the horizon of the fiducial BH, being therefore not allowed by construction.}
\label{fig:RGB_2_BH}
\end{figure}

As for the case of the single BH immersed in an external field up to the octupolar term, we have not been able to obtain a compact, closed form expression for the GB invariant on the horizon of each of the BHs. Hence, we have performed a numerical scan, illustrated in Fig.\til\ref{fig:RGB_2_BH}, fixing in each panel the value of the mass and  position parameter for the fiducial BH, and then scanning the corresponding values of the perturber. From Fig.\til\ref{fig:RGB_2_BH} it is clear that even for the two BH case, the GB invariant is always positive semi-definite on the horizon of the first (fiducial) BH\footnote{The same holds when considering the GB invariant on the horizon of the second BH.}. The figures confirm, in particular, that for smaller BH distances $\mathcal{R}_{\rm GB}$ increases on the point considered (the pole), as could be expected. 

\begin{figure}
\centering
\includegraphics[width=7cm,keepaspectratio]{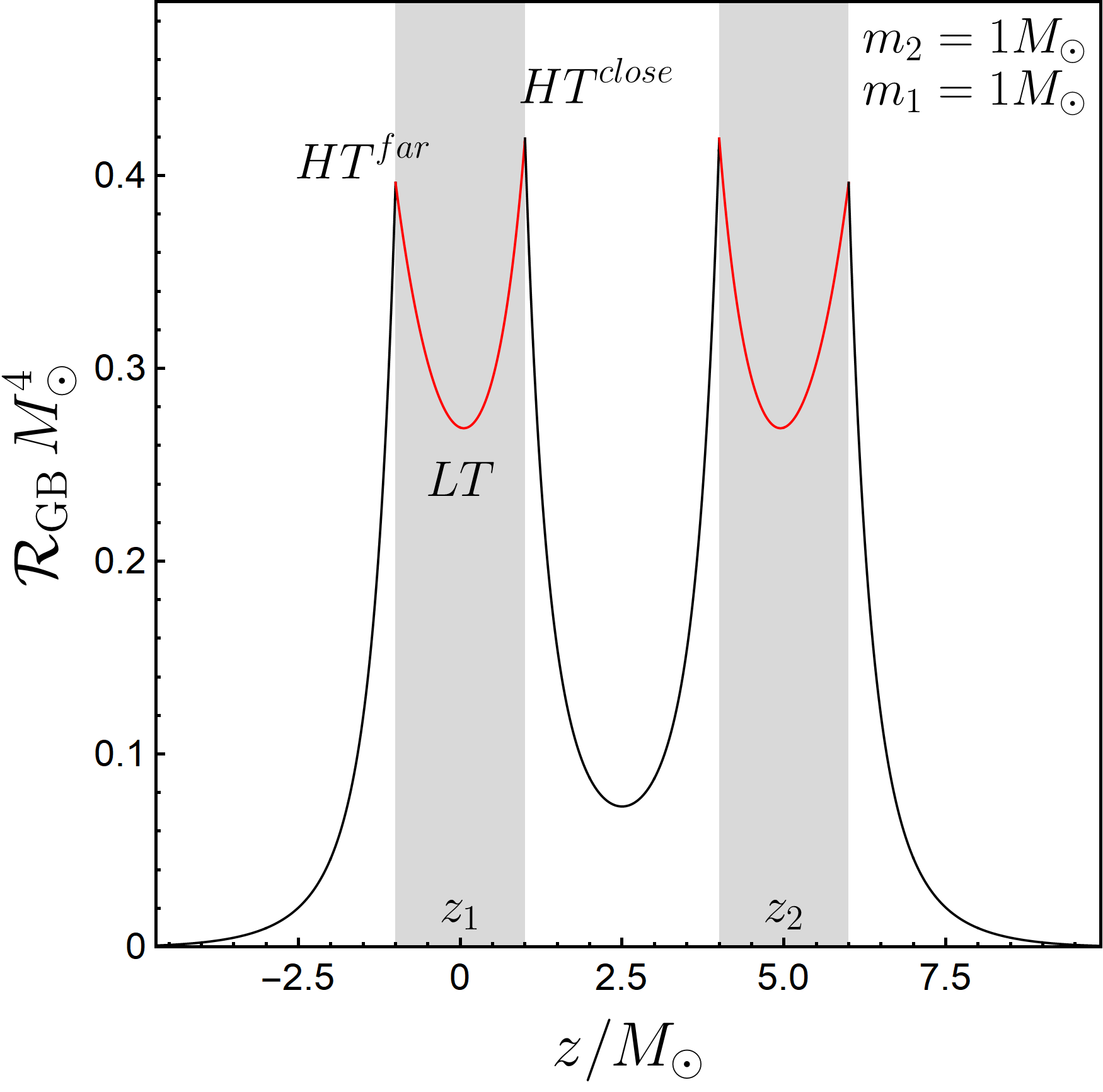} 
\includegraphics[width=7cm,keepaspectratio]{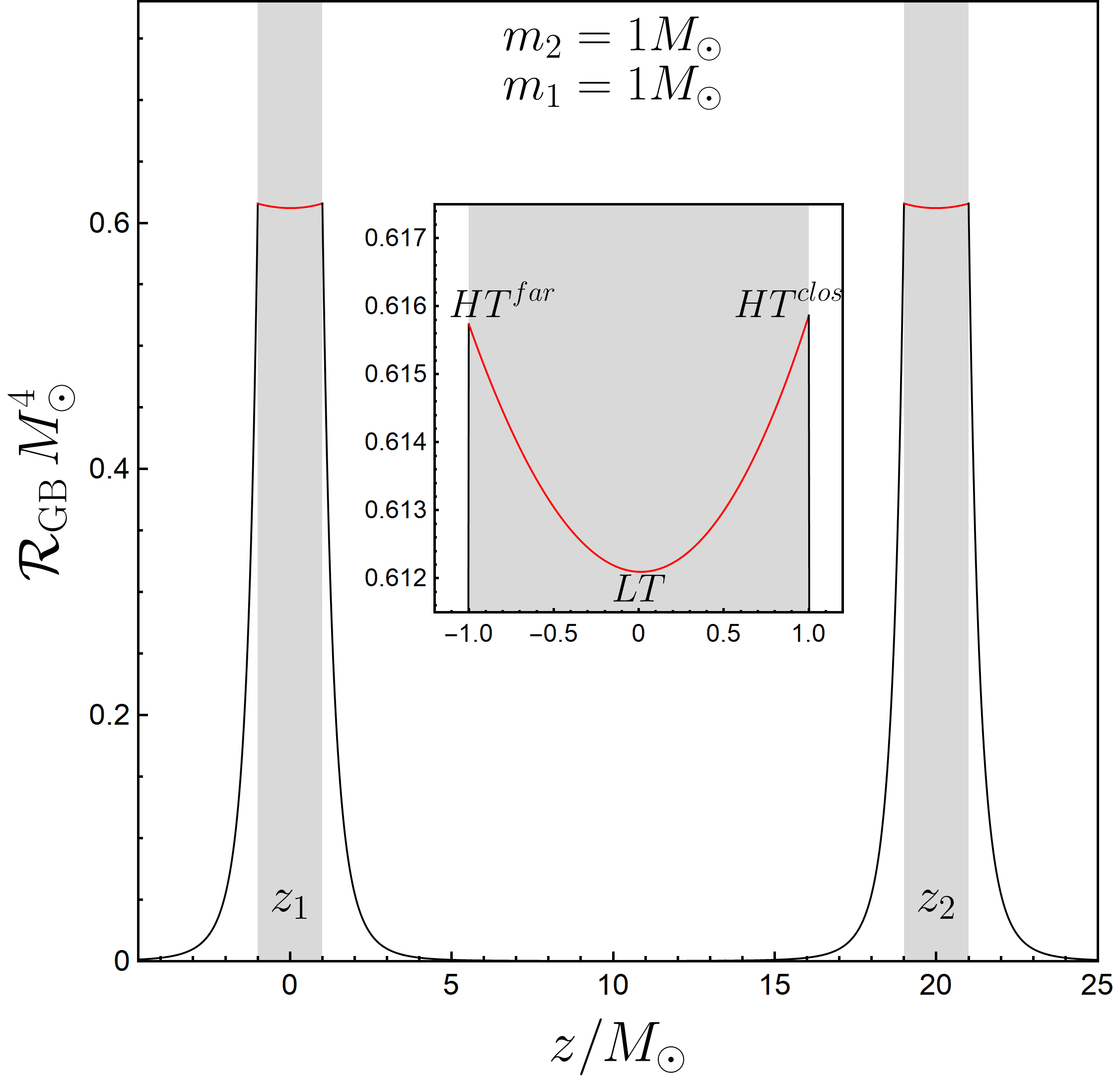} 
\caption{GB invariant along and outside the horizon of a BH in a binary ($\rho$-axis), for fixed values of  the mass and position parameters of the BHs. In both panels $m_1=m_2=M_{\odot}$ and the grey regions represents the horizons of each BH. The positions are {\bf (top)} $z_2-z_1=5M_{\odot}$, and  {\bf  (bottom)} $z_2-z_1=20M_{\odot}$. The inset zooms  $\mathcal{R}_{\rm GB}$ on the horizon of the first BH.}
\label{fig:RGB_2BH_horizon}
\end{figure}

To check the behavior of $\mathcal{R}_{\rm GB}$ on and outside the horizons, in Fig.\til\ref{fig:RGB_2BH_horizon} we show how the GB invariant varies along the $z$-axis, for two different binary configurations, none of which $\mathbb{Z}_2$-symmetric.
Focusing on the first (fiducial) BH, at $z_1=0$,  the larger the BH distance, the more the GB along the horizon of the first BH resembles the  single BH case within an external tidal environment. This is corroborated comparing the  top and bottom panels of Fig.\til\ref{fig:RGB_2BH_horizon}  where the BH distance parameter $z_2-z_1$, is varied. 
Comparing the two panels we would like to emphasise two main conclusions. Firstly, we see in the top panel that the two HT regions are \textit{asymmetric} due to the presence of the second BH. Of course, there should always be a small asymmetry, but that is not seen to leading order in perturbation theory. However, when considering a fully non-linear solution and a close by perturber, the asymmetry is manifest, whereas it becomes damped when the perturber is further away as corroborated in the inset of the bottom panel, detailing the GB invariant  along the horizon of the fiducial BH in the far-away case. 
Secondly, the  HT/LT ratio for the fiducial BH horizon reduces when the 2 BHs are displaced further away from each other (bottom panel). This is quite natural, as tidal forces decrease with distance. 
 
 %
\begin{figure}
\centering
\includegraphics[width=6.5cm,keepaspectratio]{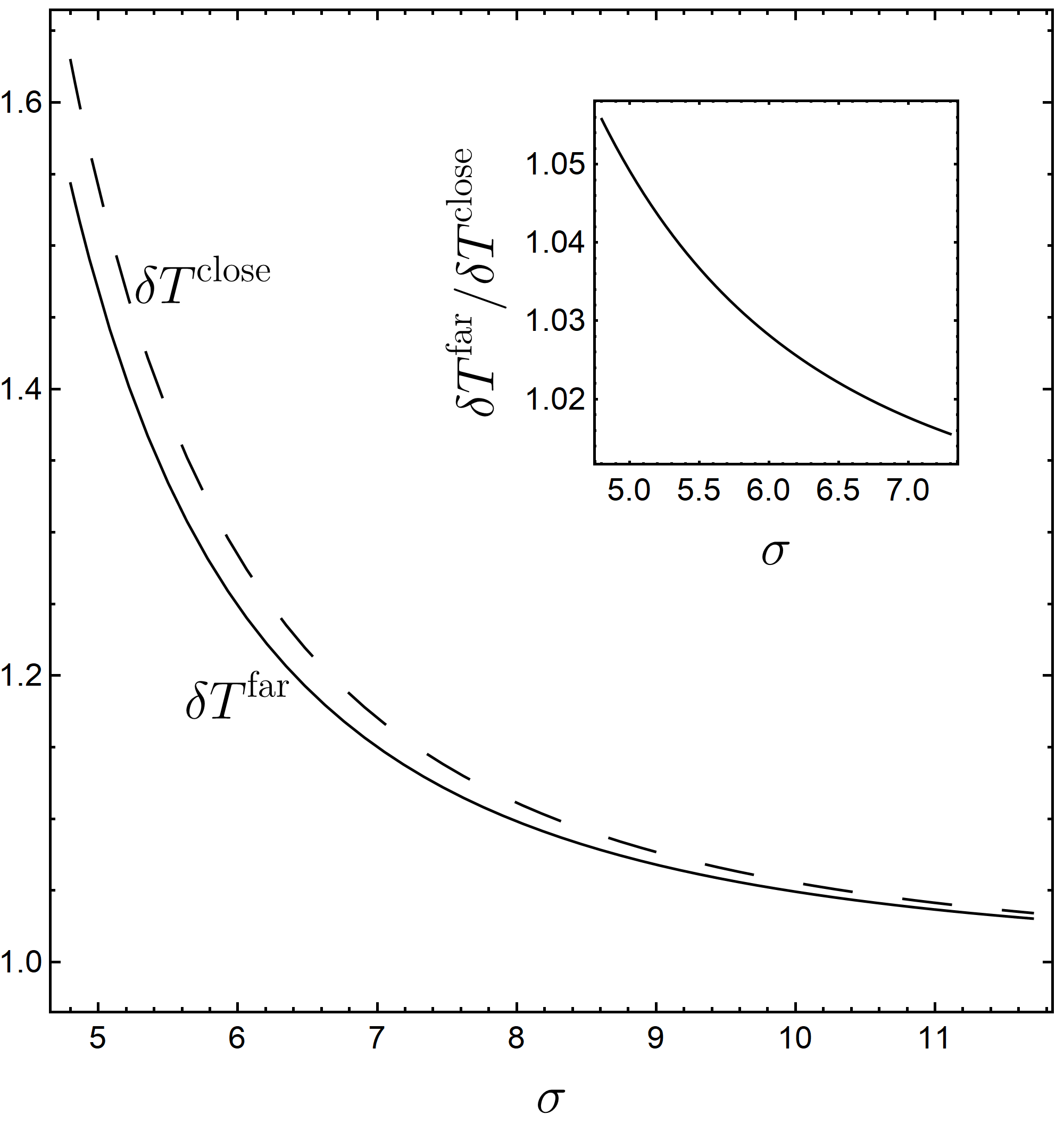} 
\caption{HT/LT  ratios computed assuming the fiducial BH at $z_1=0$, equal mass parameter binary ($m_2=m_1$) and $z=m_1$ ($\delta T^{\rm close}$) or $z=-m_1$ ($\delta T^{\rm far}$) respectively. {\bf (Inset)}  Ratio between the two HT/LT ratios for relatively large values of $\sigma=z_2/m_1$.}
\label{fig:deltaT_close_far}
\end{figure}
\begin{figure}
\centering
\includegraphics[width=8.7cm,keepaspectratio]{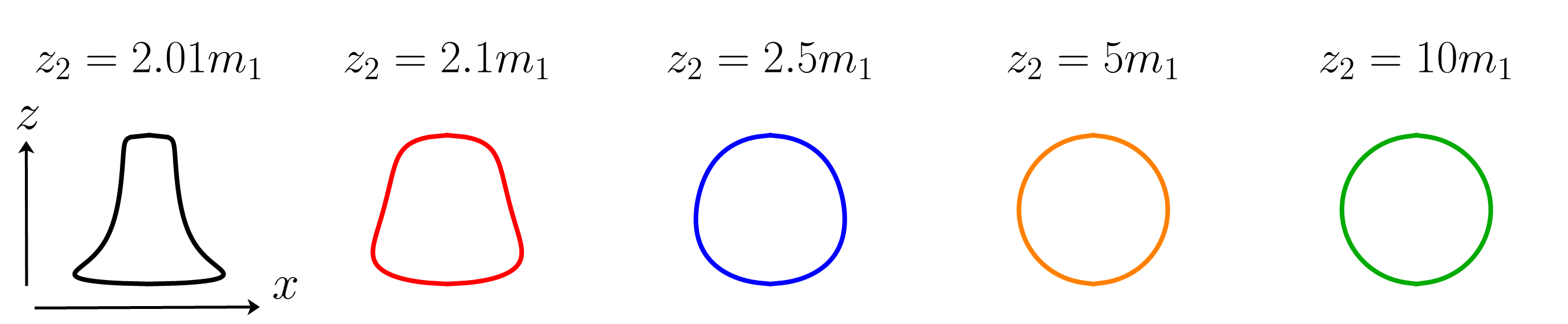} 
\caption{$x-z$ section of the  $\mathbb{E}^3$ horizons embeddings of the fiducial BH, {\it i.e.}, the one fixed in $z_1=0$, for several distances of the second BH ($z_2$). Both BH possesses the same mass parameter $m_1=m_2=1M_{\odot}$.  The specific coordinates values of each horizon embedding are scaled to show them all in the same range.}
\label{fig:embeds_2}
\end{figure}

To quantify the degree of asymmetry between the two HT regions and further understand non-linear tides when a binary is immersed in a tidal environment, we compute the tidal ratio~\eqref{deltat_2}.\footnote{ The definition of unperturbed BHs is  ambiguous for this setup. Thus, using the tidal differential  ratio $\Delta T$ defined in\til\eqref{deltat} is even more challenging than in the single BH case.} 
To simplify the calculation we fix the fiducial BH at the center ($z_1=0$) and we consider $m_2=m_1$. These  assumptions dramatically simplify the result. In fact, we are able to compute the close/far tidal ratio and provide the following expressions
\begin{align}
\delta T^{\rm close} \simeq e^{48\sigma^{-3}} \,,\,\,\,\,
\delta T^{\rm far}\simeq e^{54\sigma^{-3}}\,,
\end{align}
with $\sigma=z_2/m_2$ and accurate within $1\%$ of error for $\sigma\geq 5$, that are displayed in Fig.\til\ref{fig:deltaT_close_far}. The above fits are an accurate description of the HT/LT ratios within the provided limits, however they can be improved adding extra odd powers of the ratio $m_1/z_2$. Once again, for several BH distances $z_2$, in Fig.\til\ref{fig:embeds_2}, we construct an isometric embedding of the spatial section of the BH horizons in $\mathbb{E}^3$. Despite not being a quantitative representation of the tidal strengths, the embeddings clearly show how the asymmetry between the two HT regions is enhanced when the BHs are closer to each other, as already quantified in Fig.\til\ref{fig:deltaT_close_far}.
%

\section{Conclusion}\label{sec:Conclusions}

Some of the ground-breaking results in BH physics in the last decades concern the interaction between BHs. This is the case, {\it e.g.}, for most of the  gravitational waves transients observed by the LIGO-Virgo-KAGRA collaboration\til\cite{Abbott:2016blz,Abbott:2016nmj,Abbott:2017vtc,Abbott:2017oio,Abbott:2020tfl,LIGOScientific:2021usb}, the effects of tidal interactions on compact spacetimes\til\cite{Poisson:2011nh,Flanagan:1997sx,Pani:2015hfa}, the appearence of resonant phenomena\til\cite{Kozai:1962zz,LIDOV1962719,Palenzuela:2013hsa,Lopes:2014dba,Brito:2015oca,Bonga:2019ycj}, just to name a few.
Whereas these relate to dynamics, it is also of considerable interest to understand static or stationary BH binaries, as a window towards the non-linear interactions between compact objects in GR, which, moreover, sometimes can serve as a proxy to dynamical binaries - see $e.g.$\til\cite{Cunha:2018cof}. 

The simplest, closed-form BH binary solution in GR is (a special case of) the Majumdar–Papapetrou spacetime\til\cite{10.2307/20488481,PhysRev.72.390}, describing two charged, extremal BHs in equilibrium due to the electrostatic repulsion that counterbalances their gravitational attraction. Such extremal BHs are, however, quite special, and hardly representative of generic BHs, in particular the ones expected in astrophysical setups. In asymptotically flat vacuum GR, there has been some recent numerical constructions of equilibrium BHs with scalar hair\til\cite{Herdeiro:2023mpt,Herdeiro:2023roz}. But the lack of a closed form solution makes their analysis less transparent. On the other hand there are some non-asymptotically flat models involving external fields to balance multi-BH solutions, both analytical\til\cite{Emparan:1999au} and numerical~\cite{Dias:2023rde}. 

In this work we have used an exact Weyl-type solution of vacuum GR describing $N$ aligned, static BHs in an external, multipolar gravitational field~\cite{Astorino:2021dju,Vigano:2022hrg}, that can be interpreted as due to far away sources to probe the tidal effects of a close by, or far away companion, on a fiducial BH, in fully non-linear GR. The following are some highlights of our results.
\begin{enumerate}
\item Defining HT/LT regions as where the GB invariant  $\mathcal{R}_{\rm GB}$ (which coincides with the Kretschmann invariant in our analysis) is enhanced/suppressed, and introducing two quantitive tidal ratios, $cf.$ eqs.~\eqref{deltat} and~\eqref{deltat_2}, we observed that a (first order) perturbed BH has: (i) two symmetric HTs along the line of sight of the perturber (say, the poles) and one LT along the equator, as in Newtonian gravity; $(ii)$ $\Delta T=-2$, which means the HT is twice as strong as the LT, as in Newtonian gravity; $(iii)$ $\delta T=1+24\epsilon$ -  for instance in the Earth-Moon system, $\epsilon\simeq 10^{-35}$; on the other hand, for two Schwarzschild BHs with the same mass separated by the radius of the innermost stable circular orbit of one of them, in which case perturbation theory is clearly not valid, $\epsilon\simeq 0.005$. This means this tidal ratio is very close to unity, certainly in the weak field limit, but also beyond.
\item The exact GR solution describing a multipolar gravitational field without any BH, $cf.$ Section~\ref{subsec:External field with no black hole}, shows that for the cases analysed, including up to the octupole,  $\mathcal{R}_{\rm GB}$ is always positive, except for a zero measure set where it can  vanish. This intringuing feature is also shown to occur in a Newtonian analogue model. It would be interesting to generalize these results for an arbitrary number of multipolar components $b_n$.
\item The exact GR solution describing a single BH inside a multipolar gravitational field, including a dipole and quadrupole, $cf.$~Section~\ref{External field with one black hole}, has a special $\mathbb{Z}_2$-symmetric case for which the dipole can be set to zero and the quadrupole is arbitrary, with the solution remaining free of conical singularities. This case can be matched to the  (first order) perturbed analysis (point 1), showing that the quadrupole multipole strength is $b_2M^2=-\epsilon$. Thus, Newtonian values of $b_2M^2$ are extremely small and negative, as the Earth-Moon reference value above, but the GR solution and analysis can be extended to any sign and value of $b_2$.  The matching with perturbation theory, allowed identifying the relation between the BH mass parameter of the non-linear GR solution and the physical Komar mass, to first order in $b_2$. This allowed showing the tidal differential ratio is also $\Delta T=-2$ in this limit. On the other hand, the tidal ratio $\delta T$ could be computed for arbitrary $b_2$, showing large deviations from unity. In particular it has two critical points: it vanishes for $b_2m_1^2=1/8$ and diverges for  $b_2m_1^2=-1/2$. These critical values define three intervals, wherein this tidal ratio can be neatly associated with the 2D geometry of the spatial sections of the horizon, $cf.$ Fig.~\ref{fig:deltaT_vs_tildeQ} and~\ref{fig:embeds}.  The analysis also showed that  $\mathcal{R}_{\rm GB}$ is always non-negative, also including an octupole, both on and outside the BH horizon, and it only vanishes on a zero measure set, as in the no-BH case (point 2).
\item The exact GR solution describing two BHs inside a multipolar gravitational field, including a dipole and quadrupole, $cf.$~Section~\ref{subsec:Black_hole_binaries}, also allowed identifying a special $\mathbb{Z}_2$-symmetric case, wherein the dipole can be made to vanish, but now the quadrupole is not arbitrary.  Again, the analysis (for generic dipole and quadrupole)  showed that  $\mathcal{R}_{\rm GB}$ is always non-negative, both on and outside the BH horizon. Moreover, the analysis of the HT/LT ratio unveiled clearly asymmetric HTs, with the HT on the near and far sides relatively to the second BH differing by a few percent - $cf.$ the inset in Fig.~\ref{fig:deltaT_close_far}. In this case, the regularity condition gives no freedom to choose the multipolar strengths. For instance, for the closest configuration in Fig.~\ref{fig:RGB_2BH_horizon} (top panel), $b_1m_1^2=0.04$ and $b_2m_1^2=-0.009$. This corroborates that the value of $b_2m_1^2$  at which $\delta T$ diverges in point 3 is unreasonably high (in modulus). Thus, in this case $\delta T$ can ``only" reach values of $\simeq 1.6$, still extraordinarily large as compared to perturbation theory. 
\end{enumerate}

Let us stress that a transversal conclusion to all the cases studied is that static tides, linear or non-linear, at least within the axi-symmetric setup we have considered, cannot trigger GB$^-$ scalarization, while they seem able to enhance the GB$^+$ process. 

It would be interesting to include spin in the BHs immersed the external multipolar fields, a solution that can be constructed by an application of the inverse scattering method~\cite{Vigano:2022hrg}, to assess how spin impacts on the tidal effects described here. It would also be interesting, even if more challanging, to revisit this analysis for the numerical models of asymptotically flat, neutral, balanced binaries in\til\cite{Herdeiro:2023mpt,Herdeiro:2023roz}.

\bigskip

\subsection*{Acknowledgements}
We thank Arianna Foschi for important feedback on tidal fields around static BHs. This work is supported by the Center for Research and Development in Mathematics and Applications (CIDMA) through the Portuguese Foundation for Science and Technology (FCT - Fundaç\~{a}o para a Ci\^{e}ncia e a Tecnologia), references UIDB/04106/2020, UIDP/04106/2020. LA is supported by the University of Aveiro through a PostDoc research grant, reference BIPD/UI97/9854/2021. The authors acknowledge support from the projects PTDC/FIS-OUT/28407/2017, CERN/FISPAR/0027/2019, PTDC/FIS-AST/3041/2020 and 2022.04560.PTDC. This work has further been supported by  the  European  Union's  Horizon  2020  research  and  innovation  (RISE) programme H2020-MSCA-RISE-2017 Grant No.~FunFiCO-777740 and by the European Horizon Europe staff exchange (SE) programme HORIZON-MSCA-2021-SE-01 Grant No.~NewFunFiCO-101086251.

\bibliographystyle{apsrev4}

\bibliography{References}

\begin{thebibliography}{82}%
\makeatletter
\providecommand \@ifxundefined [1]{%
 \@ifx{#1\undefined}
}%
\providecommand \@ifnum [1]{%
 \ifnum #1\expandafter \@firstoftwo
 \else \expandafter \@secondoftwo
 \fi
}%
\providecommand \@ifx [1]{%
 \ifx #1\expandafter \@firstoftwo
 \else \expandafter \@secondoftwo
 \fi
}%
\providecommand \natexlab [1]{#1}%
\providecommand \enquote  [1]{``#1''}%
\providecommand \bibnamefont  [1]{#1}%
\providecommand \bibfnamefont [1]{#1}%
\providecommand \citenamefont [1]{#1}%
\providecommand \href@noop [0]{\@secondoftwo}%
\providecommand \href [0]{\begingroup \@sanitize@url \@href}%
\providecommand \@href[1]{\@@startlink{#1}\@@href}%
\providecommand \@@href[1]{\endgroup#1\@@endlink}%
\providecommand \@sanitize@url [0]{\catcode `\\12\catcode `\$12\catcode
  `\&12\catcode `\#12\catcode `\^12\catcode `\_12\catcode `\%12\relax}%
\providecommand \@@startlink[1]{}%
\providecommand \@@endlink[0]{}%
\providecommand \url  [0]{\begingroup\@sanitize@url \@url }%
\providecommand \@url [1]{\endgroup\@href {#1}{\urlprefix }}%
\providecommand \urlprefix  [0]{URL }%
\providecommand \Eprint [0]{\href }%
\providecommand \doibase [0]{http://dx.doi.org/}%
\providecommand \selectlanguage [0]{\@gobble}%
\providecommand \bibinfo  [0]{\@secondoftwo}%
\providecommand \bibfield  [0]{\@secondoftwo}%
\providecommand \translation [1]{[#1]}%
\providecommand \BibitemOpen [0]{}%
\providecommand \bibitemStop [0]{}%
\providecommand \bibitemNoStop [0]{.\EOS\space}%
\providecommand \EOS [0]{\spacefactor3000\relax}%
\providecommand \BibitemShut  [1]{\csname bibitem#1\endcsname}%
\let\auto@bib@innerbib\@empty
\bibitem [{\citenamefont {Newton}(1687)}]{Newton:1687eqk}%
  \BibitemOpen
  \bibfield  {author} {\bibinfo {author} {\bibfnamefont {I.}~\bibnamefont
  {Newton}}, }\href@noop {} {\emph {\bibinfo {title} {{Philosophi\ae{}
  Naturalis Principia Mathematica}}}} (\bibinfo {address} {England}, \bibinfo
  {year} {1687})\BibitemShut {NoStop}%
\bibitem [{\citenamefont {Einstein}(1915)}]{Einstein:1915ca}%
  \BibitemOpen
  \bibfield  {author} {\bibinfo {author} {\bibfnamefont {A.}~\bibnamefont
  {Einstein}}, }\href@noop {} {\bibfield  {journal} {\bibinfo  {journal} {\emph
  {Sitzungsber. Preuss. Akad. Wiss. Berlin (Math. Phys. )}} }\textbf {\bibinfo
  {volume} {1915}}, \bibinfo {pages} {844} (\bibinfo {year}
  {1915})}\BibitemShut {NoStop}%
\bibitem [{\citenamefont {Wald}(1984)}]{Wald:1984rg}%
  \BibitemOpen
  \bibfield  {author} {\bibinfo {author} {\bibfnamefont {R.M.} \bibnamefont
  {Wald}}, }\href {\doibase 10.7208/chicago/9780226870373.001.0001} {\emph
  {\bibinfo {title} {{General Relativity}}}} (\bibinfo  {publisher} {Chicago
  Univ. Pr.}, \bibinfo {address} {Chicago, USA}, \bibinfo {year}
  {1984})\BibitemShut {NoStop}%
\bibitem [{\citenamefont {d'Inverno}(1992)}]{dInverno:1992gxs}%
  \BibitemOpen
  \bibfield  {author} {\bibinfo {author} {\bibfnamefont {R.}~\bibnamefont
  {d'Inverno}}, }\href@noop {} {\emph {\bibinfo {title} {{Introducing
  Einstein's relativity}}}} (\bibinfo {year} {1992})\BibitemShut {NoStop}%
\bibitem [{\citenamefont {Lima} and \citenamefont
  {Crispino}(2020)}]{Lima:2020wcb}%
  \BibitemOpen
  \bibfield  {author} {\bibinfo {author} {\bibfnamefont {H.C.D.} \bibnamefont
  {Lima}} and \bibinfo {author} {\bibfnamefont {L.C.B.} \bibnamefont
  {Crispino}}, }\href {\doibase 10.1142/S021827182041014X} {\bibfield
  {journal} {\bibinfo  {journal} {\emph {Int. J. Mod. Phys. D}} }\textbf
  {\bibinfo {volume} {29}}, \bibinfo {pages} {2041014} (\bibinfo {year}
  {2020})}, \Eprint {http://arxiv.org/abs/2005.13029}
  {arXiv:2005.13029}\BibitemShut {NoStop}%
\bibitem [{\citenamefont {Regge} and \citenamefont
  {Wheeler}(1957)}]{Regge:1957td}%
  \BibitemOpen
  \bibfield  {author} {\bibinfo {author} {\bibfnamefont {T.}~\bibnamefont
  {Regge}} and \bibinfo {author} {\bibfnamefont {J.A.} \bibnamefont {Wheeler}},
  }\href {\doibase 10.1103/PhysRev.108.1063} {\bibfield  {journal} {\bibinfo
  {journal} {\emph {Phys.Rev.}} }\textbf {\bibinfo {volume} {108}}, \bibinfo
  {pages} {1063} (\bibinfo {year} {1957})}\BibitemShut {NoStop}%
\bibitem [{\citenamefont {Poisson}(2005)}]{Poisson:2005pi}%
  \BibitemOpen
  \bibfield  {author} {\bibinfo {author} {\bibfnamefont {E.}~\bibnamefont
  {Poisson}}, }\href {\doibase 10.1103/PhysRevLett.94.161103} {\bibfield
  {journal} {\bibinfo  {journal} {\emph {Phys. Rev. Lett.}} }\textbf {\bibinfo
  {volume} {94}}, \bibinfo {pages} {161103} (\bibinfo {year} {2005})}, \Eprint
  {http://arxiv.org/abs/gr-qc/0501032} {arXiv:gr-qc/0501032}\BibitemShut
  {NoStop}%
\bibitem [{\citenamefont {Cardoso} and \citenamefont
  {Foschi}(2021)}]{Cardoso:2021qqu}%
  \BibitemOpen
  \bibfield  {author} {\bibinfo {author} {\bibfnamefont {V.}~\bibnamefont
  {Cardoso}} and \bibinfo {author} {\bibfnamefont {A.}~\bibnamefont {Foschi}},
  }\href {\doibase 10.1103/PhysRevD.104.024004} {\bibfield  {journal} {\bibinfo
   {journal} {\emph {Phys. Rev. D}} }\textbf {\bibinfo {volume} {104}},
  \bibinfo {pages} {024004} (\bibinfo {year} {2021})}, \Eprint
  {http://arxiv.org/abs/2106.06551} {arXiv:2106.06551}\BibitemShut {NoStop}%
\bibitem [{\citenamefont {Stephani} \emph {et~al.}(2003)\citenamefont
  {Stephani}, \citenamefont {Kramer}, \citenamefont {MacCallum}, \citenamefont
  {Hoenselaers}, and \citenamefont {Herlt}}]{Stephani:2003tm}%
  \BibitemOpen
  \bibfield  {author} {\bibinfo {author} {\bibfnamefont {H.}~\bibnamefont
  {Stephani}}, \bibinfo {author} {\bibfnamefont {D.}~\bibnamefont {Kramer}},
  \bibinfo {author} {\bibfnamefont {M.A.H.} \bibnamefont {MacCallum}}, \bibinfo
  {author} {\bibfnamefont {C.}~\bibnamefont {Hoenselaers}},  and \bibinfo
  {author} {\bibfnamefont {E.}~\bibnamefont {Herlt}}, }\href {\doibase
  10.1017/CBO9780511535185} {\emph {\bibinfo {title} {{Exact solutions of
  Einstein's field equations}}}}, Cambridge Monographs on Mathematical Physics
  (\bibinfo  {publisher} {Cambridge Univ. Press}, \bibinfo {address}
  {Cambridge}, \bibinfo {year} {2003})\BibitemShut {NoStop}%
\bibitem [{\citenamefont {Doneva} \emph {et~al.}(2022)\citenamefont {Doneva},
  \citenamefont {Ramazano\u{g}lu}, \citenamefont {Silva}, \citenamefont
  {Sotiriou}, and \citenamefont {Yazadjiev}}]{Doneva:2022ewd}%
  \BibitemOpen
  \bibfield  {author} {\bibinfo {author} {\bibfnamefont {D.D.} \bibnamefont
  {Doneva}}, \bibinfo {author} {\bibfnamefont {F.M.} \bibnamefont
  {Ramazano\u{g}lu}}, \bibinfo {author} {\bibfnamefont {H.O.} \bibnamefont
  {Silva}}, \bibinfo {author} {\bibfnamefont {T.P.} \bibnamefont {Sotiriou}},
  and \bibinfo {author} {\bibfnamefont {S.S.} \bibnamefont {Yazadjiev}},
  }\href@noop {} {  (\bibinfo {year} {2022})}, \Eprint
  {http://arxiv.org/abs/2211.01766} {arXiv:2211.01766}\BibitemShut {NoStop}%
\bibitem [{\citenamefont {Herdeiro}(2022)}]{Herdeiro:2022yle}%
  \BibitemOpen
  \bibfield  {author} {\bibinfo {author} {\bibfnamefont {C.A.R.} \bibnamefont
  {Herdeiro}}, }\href@noop {} {  (\bibinfo {year} {2022})}, \Eprint
  {http://arxiv.org/abs/2204.05640} {arXiv:2204.05640}\BibitemShut {NoStop}%
\bibitem [{\citenamefont {Silva} \emph {et~al.}(2018)\citenamefont {Silva},
  \citenamefont {Sakstein}, \citenamefont {Gualtieri}, \citenamefont
  {Sotiriou}, and \citenamefont {Berti}}]{Silva:2017uqg}%
  \BibitemOpen
  \bibfield  {author} {\bibinfo {author} {\bibfnamefont {H.O.} \bibnamefont
  {Silva}}, \bibinfo {author} {\bibfnamefont {J.}~\bibnamefont {Sakstein}},
  \bibinfo {author} {\bibfnamefont {L.}~\bibnamefont {Gualtieri}}, \bibinfo
  {author} {\bibfnamefont {T.P.} \bibnamefont {Sotiriou}},  and \bibinfo
  {author} {\bibfnamefont {E.}~\bibnamefont {Berti}}, }\href {\doibase
  10.1103/PhysRevLett.120.131104} {\bibfield  {journal} {\bibinfo  {journal}
  {\emph {Phys. Rev. Lett.}} }\textbf {\bibinfo {volume} {120}}, \bibinfo
  {pages} {131104} (\bibinfo {year} {2018})}, \Eprint
  {http://arxiv.org/abs/1711.02080} {arXiv:1711.02080}\BibitemShut {NoStop}%
\bibitem [{\citenamefont {Doneva} and \citenamefont
  {Yazadjiev}(2018{\natexlab{a}})}]{Doneva:2017bvd}%
  \BibitemOpen
  \bibfield  {author} {\bibinfo {author} {\bibfnamefont {D.D.} \bibnamefont
  {Doneva}} and \bibinfo {author} {\bibfnamefont {S.S.} \bibnamefont
  {Yazadjiev}}, }\href {\doibase 10.1103/PhysRevLett.120.131103} {\bibfield
  {journal} {\bibinfo  {journal} {\emph {Phys. Rev. Lett.}} }\textbf {\bibinfo
  {volume} {120}}, \bibinfo {pages} {131103} (\bibinfo {year}
  {2018}{\natexlab{a}})}, \Eprint {http://arxiv.org/abs/1711.01187}
  {arXiv:1711.01187}\BibitemShut {NoStop}%
\bibitem [{\citenamefont {Witek} \emph {et~al.}(2019)\citenamefont {Witek},
  \citenamefont {Gualtieri}, \citenamefont {Pani}, and \citenamefont
  {Sotiriou}}]{Witek:2018dmd}%
  \BibitemOpen
  \bibfield  {author} {\bibinfo {author} {\bibfnamefont {H.}~\bibnamefont
  {Witek}}, \bibinfo {author} {\bibfnamefont {L.}~\bibnamefont {Gualtieri}},
  \bibinfo {author} {\bibfnamefont {P.}~\bibnamefont {Pani}},  and \bibinfo
  {author} {\bibfnamefont {T.P.} \bibnamefont {Sotiriou}}, }\href {\doibase
  10.1103/PhysRevD.99.064035} {\bibfield  {journal} {\bibinfo  {journal} {\emph
  {Phys. Rev. D}} }\textbf {\bibinfo {volume} {99}}, \bibinfo {pages} {064035}
  (\bibinfo {year} {2019})}, \Eprint {http://arxiv.org/abs/1810.05177}
  {arXiv:1810.05177}\BibitemShut {NoStop}%
\bibitem [{\citenamefont {Silva} \emph {et~al.}(2019)\citenamefont {Silva},
  \citenamefont {Macedo}, \citenamefont {Sotiriou}, \citenamefont {Gualtieri},
  \citenamefont {Sakstein}, and \citenamefont {Berti}}]{Silva:2018qhn}%
  \BibitemOpen
  \bibfield  {author} {\bibinfo {author} {\bibfnamefont {H.O.} \bibnamefont
  {Silva}}, \bibinfo {author} {\bibfnamefont {C.F.} \bibnamefont {Macedo}},
  \bibinfo {author} {\bibfnamefont {T.P.} \bibnamefont {Sotiriou}}, \bibinfo
  {author} {\bibfnamefont {L.}~\bibnamefont {Gualtieri}}, \bibinfo {author}
  {\bibfnamefont {J.}~\bibnamefont {Sakstein}},  and \bibinfo {author}
  {\bibfnamefont {E.}~\bibnamefont {Berti}}, }\href {\doibase
  10.1103/PhysRevD.99.064011} {\bibfield  {journal} {\bibinfo  {journal} {\emph
  {Phys. Rev. D}} }\textbf {\bibinfo {volume} {99}}, \bibinfo {pages} {064011}
  (\bibinfo {year} {2019})}, \Eprint {http://arxiv.org/abs/1812.05590}
  {arXiv:1812.05590}\BibitemShut {NoStop}%
\bibitem [{\citenamefont {Minamitsuji} and \citenamefont
  {Ikeda}(2019{\natexlab{a}})}]{Minamitsuji:2018xde}%
  \BibitemOpen
  \bibfield  {author} {\bibinfo {author} {\bibfnamefont {M.}~\bibnamefont
  {Minamitsuji}} and \bibinfo {author} {\bibfnamefont {T.}~\bibnamefont
  {Ikeda}}, }\href {\doibase 10.1103/PhysRevD.99.044017} {\bibfield  {journal}
  {\bibinfo  {journal} {\emph {Phys. Rev. D}} }\textbf {\bibinfo {volume}
  {99}}, \bibinfo {pages} {044017} (\bibinfo {year} {2019}{\natexlab{a}})},
  \Eprint {http://arxiv.org/abs/1812.03551} {arXiv:1812.03551}\BibitemShut
  {NoStop}%
\bibitem [{\citenamefont {Doneva} \emph {et~al.}(2019)\citenamefont {Doneva},
  \citenamefont {Staykov}, and \citenamefont {Yazadjiev}}]{Doneva:2019vuh}%
  \BibitemOpen
  \bibfield  {author} {\bibinfo {author} {\bibfnamefont {D.D.} \bibnamefont
  {Doneva}}, \bibinfo {author} {\bibfnamefont {K.V.} \bibnamefont {Staykov}},
  and \bibinfo {author} {\bibfnamefont {S.S.} \bibnamefont {Yazadjiev}}, }\href
  {\doibase 10.1103/PhysRevD.99.104045} {\bibfield  {journal} {\bibinfo
  {journal} {\emph {Phys. Rev. D}} }\textbf {\bibinfo {volume} {99}}, \bibinfo
  {pages} {104045} (\bibinfo {year} {2019})}, \Eprint
  {http://arxiv.org/abs/1903.08119} {arXiv:1903.08119}\BibitemShut {NoStop}%
\bibitem [{\citenamefont {Fernandes} \emph {et~al.}(2019)\citenamefont
  {Fernandes}, \citenamefont {Herdeiro}, \citenamefont {Pombo}, \citenamefont
  {Radu}, and \citenamefont {Sanchis-Gual}}]{Fernandes:2019rez}%
  \BibitemOpen
  \bibfield  {author} {\bibinfo {author} {\bibfnamefont {P.G.} \bibnamefont
  {Fernandes}}, \bibinfo {author} {\bibfnamefont {C.A.} \bibnamefont
  {Herdeiro}}, \bibinfo {author} {\bibfnamefont {A.M.} \bibnamefont {Pombo}},
  \bibinfo {author} {\bibfnamefont {E.}~\bibnamefont {Radu}},  and \bibinfo
  {author} {\bibfnamefont {N.}~\bibnamefont {Sanchis-Gual}}, }\href {\doibase
  10.1088/1361-6382/ab23a1} {\bibfield  {journal} {\bibinfo  {journal} {\emph
  {Class. Quant. Grav.}} }\textbf {\bibinfo {volume} {36}}, \bibinfo {pages}
  {134002} (\bibinfo {year} {2019})}, \bibinfo {note} {[Erratum:
  Class.Quant.Grav. 37, 049501 (2020)]}, \Eprint
  {http://arxiv.org/abs/1902.05079} {arXiv:1902.05079}\BibitemShut {NoStop}%
\bibitem [{\citenamefont {Minamitsuji} and \citenamefont
  {Ikeda}(2019{\natexlab{b}})}]{Minamitsuji:2019iwp}%
  \BibitemOpen
  \bibfield  {author} {\bibinfo {author} {\bibfnamefont {M.}~\bibnamefont
  {Minamitsuji}} and \bibinfo {author} {\bibfnamefont {T.}~\bibnamefont
  {Ikeda}}, }\href {\doibase 10.1103/PhysRevD.99.104069} {\bibfield  {journal}
  {\bibinfo  {journal} {\emph {Phys. Rev. D}} }\textbf {\bibinfo {volume}
  {99}}, \bibinfo {pages} {104069} (\bibinfo {year} {2019}{\natexlab{b}})},
  \Eprint {http://arxiv.org/abs/1904.06572} {arXiv:1904.06572}\BibitemShut
  {NoStop}%
\bibitem [{\citenamefont {Cunha} \emph {et~al.}(2019)\citenamefont {Cunha},
  \citenamefont {Herdeiro}, and \citenamefont {Radu}}]{Cunha:2019dwb}%
  \BibitemOpen
  \bibfield  {author} {\bibinfo {author} {\bibfnamefont {P.V.} \bibnamefont
  {Cunha}}, \bibinfo {author} {\bibfnamefont {C.A.} \bibnamefont {Herdeiro}},
  and \bibinfo {author} {\bibfnamefont {E.}~\bibnamefont {Radu}}, }\href
  {\doibase 10.1103/PhysRevLett.123.011101} {\bibfield  {journal} {\bibinfo
  {journal} {\emph {Phys. Rev. Lett.}} }\textbf {\bibinfo {volume} {123}},
  \bibinfo {pages} {011101} (\bibinfo {year} {2019})}, \Eprint
  {http://arxiv.org/abs/1904.09997} {arXiv:1904.09997}\BibitemShut {NoStop}%
\bibitem [{\citenamefont {Andreou} \emph {et~al.}(2019)\citenamefont {Andreou},
  \citenamefont {Franchini}, \citenamefont {Ventagli}, and \citenamefont
  {Sotiriou}}]{Andreou:2019ikc}%
  \BibitemOpen
  \bibfield  {author} {\bibinfo {author} {\bibfnamefont {N.}~\bibnamefont
  {Andreou}}, \bibinfo {author} {\bibfnamefont {N.}~\bibnamefont {Franchini}},
  \bibinfo {author} {\bibfnamefont {G.}~\bibnamefont {Ventagli}},  and \bibinfo
  {author} {\bibfnamefont {T.P.} \bibnamefont {Sotiriou}}, }\href {\doibase
  10.1103/PhysRevD.99.124022} {\bibfield  {journal} {\bibinfo  {journal} {\emph
  {Phys. Rev. D}} }\textbf {\bibinfo {volume} {99}}, \bibinfo {pages} {124022}
  (\bibinfo {year} {2019})}, \bibinfo {note} {[Erratum: Phys.Rev.D 101, 109903
  (2020)]}, \Eprint {http://arxiv.org/abs/1904.06365}
  {arXiv:1904.06365}\BibitemShut {NoStop}%
\bibitem [{\citenamefont {Ikeda} \emph {et~al.}(2019)\citenamefont {Ikeda},
  \citenamefont {Nakamura}, and \citenamefont {Minamitsuji}}]{Ikeda:2019okp}%
  \BibitemOpen
  \bibfield  {author} {\bibinfo {author} {\bibfnamefont {T.}~\bibnamefont
  {Ikeda}}, \bibinfo {author} {\bibfnamefont {T.}~\bibnamefont {Nakamura}},
  and \bibinfo {author} {\bibfnamefont {M.}~\bibnamefont {Minamitsuji}}, }\href
  {\doibase 10.1103/PhysRevD.100.104014} {\bibfield  {journal} {\bibinfo
  {journal} {\emph {Phys. Rev. D}} }\textbf {\bibinfo {volume} {100}}, \bibinfo
  {pages} {104014} (\bibinfo {year} {2019})}, \Eprint
  {http://arxiv.org/abs/1908.09394} {arXiv:1908.09394}\BibitemShut {NoStop}%
\bibitem [{\citenamefont {Ramazano\u{g}lu}(2017)}]{Ramazanoglu:2017xbl}%
  \BibitemOpen
  \bibfield  {author} {\bibinfo {author} {\bibfnamefont {F.M.} \bibnamefont
  {Ramazano\u{g}lu}}, }\href {\doibase 10.1103/PhysRevD.96.064009} {\bibfield
  {journal} {\bibinfo  {journal} {\emph {Phys. Rev. D}} }\textbf {\bibinfo
  {volume} {96}}, \bibinfo {pages} {064009} (\bibinfo {year} {2017})}, \Eprint
  {http://arxiv.org/abs/1706.01056} {arXiv:1706.01056}\BibitemShut {NoStop}%
\bibitem [{\citenamefont {Doneva} and \citenamefont
  {Yazadjiev}(2018{\natexlab{b}})}]{Doneva:2017duq}%
  \BibitemOpen
  \bibfield  {author} {\bibinfo {author} {\bibfnamefont {D.D.} \bibnamefont
  {Doneva}} and \bibinfo {author} {\bibfnamefont {S.S.} \bibnamefont
  {Yazadjiev}}, }\href {\doibase 10.1088/1475-7516/2018/04/011} {\bibfield
  {journal} {\bibinfo  {journal} {\emph {JCAP}} }\textbf {\bibinfo {volume}
  {04}}, \bibinfo {pages} {011} (\bibinfo {year} {2018}{\natexlab{b}})},
  \Eprint {http://arxiv.org/abs/1712.03715} {arXiv:1712.03715}\BibitemShut
  {NoStop}%
\bibitem [{\citenamefont {Annulli} \emph {et~al.}(2019)\citenamefont {Annulli},
  \citenamefont {Cardoso}, and \citenamefont {Gualtieri}}]{Annulli:2019fzq}%
  \BibitemOpen
  \bibfield  {author} {\bibinfo {author} {\bibfnamefont {L.}~\bibnamefont
  {Annulli}}, \bibinfo {author} {\bibfnamefont {V.}~\bibnamefont {Cardoso}},
  and \bibinfo {author} {\bibfnamefont {L.}~\bibnamefont {Gualtieri}}, }\href
  {\doibase 10.1103/PhysRevD.99.044038} {\bibfield  {journal} {\bibinfo
  {journal} {\emph {Phys. Rev. D}} }\textbf {\bibinfo {volume} {99}}, \bibinfo
  {pages} {044038} (\bibinfo {year} {2019})}, \Eprint
  {http://arxiv.org/abs/1901.02461} {arXiv:1901.02461}\BibitemShut {NoStop}%
\bibitem [{\citenamefont {Kase} \emph {et~al.}(2020)\citenamefont {Kase},
  \citenamefont {Minamitsuji}, and \citenamefont {Tsujikawa}}]{Kase:2020yhw}%
  \BibitemOpen
  \bibfield  {author} {\bibinfo {author} {\bibfnamefont {R.}~\bibnamefont
  {Kase}}, \bibinfo {author} {\bibfnamefont {M.}~\bibnamefont {Minamitsuji}},
  and \bibinfo {author} {\bibfnamefont {S.}~\bibnamefont {Tsujikawa}}, }\href
  {\doibase 10.1103/PhysRevD.102.024067} {\bibfield  {journal} {\bibinfo
  {journal} {\emph {Phys. Rev. D}} }\textbf {\bibinfo {volume} {102}}, \bibinfo
  {pages} {024067} (\bibinfo {year} {2020})}, \Eprint
  {http://arxiv.org/abs/2001.10701} {arXiv:2001.10701}\BibitemShut {NoStop}%
\bibitem [{\citenamefont {Ramazano\u{g}lu}(2019)}]{Ramazanoglu:2019gbz}%
  \BibitemOpen
  \bibfield  {author} {\bibinfo {author} {\bibfnamefont {F.M.} \bibnamefont
  {Ramazano\u{g}lu}}, }\href {\doibase 10.1103/PhysRevD.99.084015} {\bibfield
  {journal} {\bibinfo  {journal} {\emph {Phys. Rev. D}} }\textbf {\bibinfo
  {volume} {99}}, \bibinfo {pages} {084015} (\bibinfo {year} {2019})}, \Eprint
  {http://arxiv.org/abs/1901.10009} {arXiv:1901.10009}\BibitemShut {NoStop}%
\bibitem [{\citenamefont
  {Ramazano\u{g}lu}(2018{\natexlab{a}})}]{Ramazanoglu:2017yun}%
  \BibitemOpen
  \bibfield  {author} {\bibinfo {author} {\bibfnamefont {F.M.} \bibnamefont
  {Ramazano\u{g}lu}}, }\href {\doibase 10.1103/PhysRevD.97.024008} {\bibfield
  {journal} {\bibinfo  {journal} {\emph {Phys. Rev. D}} }\textbf {\bibinfo
  {volume} {97}}, \bibinfo {pages} {024008} (\bibinfo {year}
  {2018}{\natexlab{a}})}, \bibinfo {note} {[Erratum: Phys.Rev.D 99, 069905
  (2019)]}, \Eprint {http://arxiv.org/abs/1710.00863}
  {arXiv:1710.00863}\BibitemShut {NoStop}%
\bibitem [{\citenamefont
  {Ramazano\u{g}lu}(2018{\natexlab{b}})}]{Ramazanoglu:2018hwk}%
  \BibitemOpen
  \bibfield  {author} {\bibinfo {author} {\bibfnamefont {F.M.} \bibnamefont
  {Ramazano\u{g}lu}}, }\href {\doibase 10.1103/PhysRevD.98.044011} {\bibfield
  {journal} {\bibinfo  {journal} {\emph {Phys. Rev. D}} }\textbf {\bibinfo
  {volume} {98}}, \bibinfo {pages} {044011} (\bibinfo {year}
  {2018}{\natexlab{b}})}, \bibinfo {note} {[Erratum: Phys.Rev.D 100, 029903
  (2019)]}, \Eprint {http://arxiv.org/abs/1804.00594}
  {arXiv:1804.00594}\BibitemShut {NoStop}%
\bibitem [{\citenamefont {Minamitsuji}(2020)}]{Minamitsuji:2020hpl}%
  \BibitemOpen
  \bibfield  {author} {\bibinfo {author} {\bibfnamefont {M.}~\bibnamefont
  {Minamitsuji}}, }\href {\doibase 10.1103/PhysRevD.102.044048} {\bibfield
  {journal} {\bibinfo  {journal} {\emph {Phys. Rev. D}} }\textbf {\bibinfo
  {volume} {102}}, \bibinfo {pages} {044048} (\bibinfo {year} {2020})}, \Eprint
  {http://arxiv.org/abs/2008.12758} {arXiv:2008.12758}\BibitemShut {NoStop}%
\bibitem [{\citenamefont {Antoniou} \emph {et~al.}(2018)\citenamefont
  {Antoniou}, \citenamefont {Bakopoulos}, and \citenamefont
  {Kanti}}]{Antoniou:2017acq}%
  \BibitemOpen
  \bibfield  {author} {\bibinfo {author} {\bibfnamefont {G.}~\bibnamefont
  {Antoniou}}, \bibinfo {author} {\bibfnamefont {A.}~\bibnamefont
  {Bakopoulos}},  and \bibinfo {author} {\bibfnamefont {P.}~\bibnamefont
  {Kanti}}, }\href {\doibase 10.1103/PhysRevLett.120.131102} {\bibfield
  {journal} {\bibinfo  {journal} {\emph {Phys. Rev. Lett.}} }\textbf {\bibinfo
  {volume} {120}}, \bibinfo {pages} {131102} (\bibinfo {year} {2018})}, \Eprint
  {http://arxiv.org/abs/1711.03390} {arXiv:1711.03390}\BibitemShut {NoStop}%
\bibitem [{\citenamefont {Annulli}(2021)}]{Annulli:2021lmn}%
  \BibitemOpen
  \bibfield  {author} {\bibinfo {author} {\bibfnamefont {L.}~\bibnamefont
  {Annulli}}, }\href {\doibase 10.1103/PhysRevD.104.124028} {\bibfield
  {journal} {\bibinfo  {journal} {\emph {Phys. Rev. D}} }\textbf {\bibinfo
  {volume} {104}}, \bibinfo {pages} {124028} (\bibinfo {year} {2021})}, \Eprint
  {http://arxiv.org/abs/2105.08728} {arXiv:2105.08728}\BibitemShut {NoStop}%
\bibitem [{\citenamefont {Herdeiro} \emph
  {et~al.}(2021{\natexlab{a}})\citenamefont {Herdeiro}, \citenamefont {Pombo},
  and \citenamefont {Radu}}]{Herdeiro:2021vjo}%
  \BibitemOpen
  \bibfield  {author} {\bibinfo {author} {\bibfnamefont {C.A.R.} \bibnamefont
  {Herdeiro}}, \bibinfo {author} {\bibfnamefont {A.M.} \bibnamefont {Pombo}},
  and \bibinfo {author} {\bibfnamefont {E.}~\bibnamefont {Radu}}, }\href
  {\doibase 10.3390/universe7120483} {\bibfield  {journal} {\bibinfo  {journal}
  {\emph {Universe}} }\textbf {\bibinfo {volume} {7}}, \bibinfo {pages} {483}
  (\bibinfo {year} {2021}{\natexlab{a}})}, \Eprint
  {http://arxiv.org/abs/2111.06442} {arXiv:2111.06442}\BibitemShut {NoStop}%
\bibitem [{\citenamefont {Dima} \emph {et~al.}(2020)\citenamefont {Dima},
  \citenamefont {Barausse}, \citenamefont {Franchini}, and \citenamefont
  {Sotiriou}}]{Dima:2020yac}%
  \BibitemOpen
  \bibfield  {author} {\bibinfo {author} {\bibfnamefont {A.}~\bibnamefont
  {Dima}}, \bibinfo {author} {\bibfnamefont {E.}~\bibnamefont {Barausse}},
  \bibinfo {author} {\bibfnamefont {N.}~\bibnamefont {Franchini}},  and
  \bibinfo {author} {\bibfnamefont {T.P.} \bibnamefont {Sotiriou}}, }\href
  {\doibase 10.1103/PhysRevLett.125.231101} {\bibfield  {journal} {\bibinfo
  {journal} {\emph {Phys. Rev. Lett.}} }\textbf {\bibinfo {volume} {125}},
  \bibinfo {pages} {231101} (\bibinfo {year} {2020})}, \Eprint
  {http://arxiv.org/abs/2006.03095} {arXiv:2006.03095}\BibitemShut {NoStop}%
\bibitem [{\citenamefont {Herdeiro} \emph
  {et~al.}(2021{\natexlab{b}})\citenamefont {Herdeiro}, \citenamefont {Radu},
  \citenamefont {Silva}, \citenamefont {Sotiriou}, and \citenamefont
  {Yunes}}]{Herdeiro:2020wei}%
  \BibitemOpen
  \bibfield  {author} {\bibinfo {author} {\bibfnamefont {C.A.R.} \bibnamefont
  {Herdeiro}}, \bibinfo {author} {\bibfnamefont {E.}~\bibnamefont {Radu}},
  \bibinfo {author} {\bibfnamefont {H.O.} \bibnamefont {Silva}}, \bibinfo
  {author} {\bibfnamefont {T.P.} \bibnamefont {Sotiriou}},  and \bibinfo
  {author} {\bibfnamefont {N.}~\bibnamefont {Yunes}}, }\href {\doibase
  10.1103/PhysRevLett.126.011103} {\bibfield  {journal} {\bibinfo  {journal}
  {\emph {Phys. Rev. Lett.}} }\textbf {\bibinfo {volume} {126}}, \bibinfo
  {pages} {011103} (\bibinfo {year} {2021}{\natexlab{b}})}, \Eprint
  {http://arxiv.org/abs/2009.03904} {arXiv:2009.03904}\BibitemShut {NoStop}%
\bibitem [{\citenamefont {Berti} \emph {et~al.}(2021)\citenamefont {Berti},
  \citenamefont {Collodel}, \citenamefont {Kleihaus}, and \citenamefont
  {Kunz}}]{Berti:2020kgk}%
  \BibitemOpen
  \bibfield  {author} {\bibinfo {author} {\bibfnamefont {E.}~\bibnamefont
  {Berti}}, \bibinfo {author} {\bibfnamefont {L.G.} \bibnamefont {Collodel}},
  \bibinfo {author} {\bibfnamefont {B.}~\bibnamefont {Kleihaus}},  and \bibinfo
  {author} {\bibfnamefont {J.}~\bibnamefont {Kunz}}, }\href {\doibase
  10.1103/PhysRevLett.126.011104} {\bibfield  {journal} {\bibinfo  {journal}
  {\emph {Phys. Rev. Lett.}} }\textbf {\bibinfo {volume} {126}}, \bibinfo
  {pages} {011104} (\bibinfo {year} {2021})}, \Eprint
  {http://arxiv.org/abs/2009.03905} {arXiv:2009.03905}\BibitemShut {NoStop}%
\bibitem [{\citenamefont {Herdeiro} \emph {et~al.}(2018)\citenamefont
  {Herdeiro}, \citenamefont {Radu}, \citenamefont {Sanchis-Gual}, and
  \citenamefont {Font}}]{Herdeiro:2018wub}%
  \BibitemOpen
  \bibfield  {author} {\bibinfo {author} {\bibfnamefont {C.A.R.} \bibnamefont
  {Herdeiro}}, \bibinfo {author} {\bibfnamefont {E.}~\bibnamefont {Radu}},
  \bibinfo {author} {\bibfnamefont {N.}~\bibnamefont {Sanchis-Gual}},  and
  \bibinfo {author} {\bibfnamefont {J.A.} \bibnamefont {Font}}, }\href
  {\doibase 10.1103/PhysRevLett.121.101102} {\bibfield  {journal} {\bibinfo
  {journal} {\emph {Phys. Rev. Lett.}} }\textbf {\bibinfo {volume} {121}},
  \bibinfo {pages} {101102} (\bibinfo {year} {2018})}, \Eprint
  {http://arxiv.org/abs/1806.05190} {arXiv:1806.05190}\BibitemShut {NoStop}%
\bibitem [{\citenamefont {Brihaye} \emph {et~al.}(2021)\citenamefont {Brihaye},
  \citenamefont {Capobianco}, and \citenamefont {Hartmann}}]{Brihaye:2021jop}%
  \BibitemOpen
  \bibfield  {author} {\bibinfo {author} {\bibfnamefont {Y.}~\bibnamefont
  {Brihaye}}, \bibinfo {author} {\bibfnamefont {R.}~\bibnamefont {Capobianco}},
   and \bibinfo {author} {\bibfnamefont {B.}~\bibnamefont {Hartmann}}, }\href
  {\doibase 10.1103/PhysRevD.103.124020} {\bibfield  {journal} {\bibinfo
  {journal} {\emph {Phys. Rev. D}} }\textbf {\bibinfo {volume} {103}}, \bibinfo
  {pages} {124020} (\bibinfo {year} {2021})}, \Eprint
  {http://arxiv.org/abs/2103.09307} {arXiv:2103.09307}\BibitemShut {NoStop}%
\bibitem [{\citenamefont {Annulli} \emph {et~al.}(2022)\citenamefont {Annulli},
  \citenamefont {Herdeiro}, and \citenamefont {Radu}}]{Annulli:2022ivr}%
  \BibitemOpen
  \bibfield  {author} {\bibinfo {author} {\bibfnamefont {L.}~\bibnamefont
  {Annulli}}, \bibinfo {author} {\bibfnamefont {C.A.R.} \bibnamefont
  {Herdeiro}},  and \bibinfo {author} {\bibfnamefont {E.}~\bibnamefont {Radu}},
  }\href {\doibase 10.1016/j.physletb.2022.137227} {\bibfield  {journal}
  {\bibinfo  {journal} {\emph {Phys. Lett. B}} }\textbf {\bibinfo {volume}
  {832}}, \bibinfo {pages} {137227} (\bibinfo {year} {2022})}, \Eprint
  {http://arxiv.org/abs/2203.13267} {arXiv:2203.13267}\BibitemShut {NoStop}%
\bibitem [{\citenamefont {Gardner} \emph {et~al.}(1967)\citenamefont {Gardner},
  \citenamefont {Greene}, \citenamefont {Kruskal}, and \citenamefont
  {Miura}}]{PhysRevLett.19.1095}%
  \BibitemOpen
  \bibfield  {author} {\bibinfo {author} {\bibfnamefont {C.S.} \bibnamefont
  {Gardner}}, \bibinfo {author} {\bibfnamefont {J.M.} \bibnamefont {Greene}},
  \bibinfo {author} {\bibfnamefont {M.D.} \bibnamefont {Kruskal}},  and
  \bibinfo {author} {\bibfnamefont {R.M.} \bibnamefont {Miura}}, }\href
  {\doibase 10.1103/PhysRevLett.19.1095} {\bibfield  {journal} {\bibinfo
  {journal} {\emph {Phys. Rev. Lett.}} }\textbf {\bibinfo {volume} {19}},
  \bibinfo {pages} {1095} (\bibinfo {year} {1967})}\BibitemShut {NoStop}%
\bibitem [{\citenamefont {Ernst}(1968{\natexlab{a}})}]{PhysRev.167.1175}%
  \BibitemOpen
  \bibfield  {author} {\bibinfo {author} {\bibfnamefont {F.J.} \bibnamefont
  {Ernst}}, }\href {\doibase 10.1103/PhysRev.167.1175} {\bibfield  {journal}
  {\bibinfo  {journal} {\emph {Phys. Rev.}} }\textbf {\bibinfo {volume} {167}},
  \bibinfo {pages} {1175} (\bibinfo {year} {1968}{\natexlab{a}})}\BibitemShut
  {NoStop}%
\bibitem [{\citenamefont {Ernst}(1968{\natexlab{b}})}]{PhysRev.168.1415}%
  \BibitemOpen
  \bibfield  {author} {\bibinfo {author} {\bibfnamefont {F.J.} \bibnamefont
  {Ernst}}, }\href {\doibase 10.1103/PhysRev.168.1415} {\bibfield  {journal}
  {\bibinfo  {journal} {\emph {Phys. Rev.}} }\textbf {\bibinfo {volume} {168}},
  \bibinfo {pages} {1415} (\bibinfo {year} {1968}{\natexlab{b}})}\BibitemShut
  {NoStop}%
\bibitem [{\citenamefont {Alekseev}(2010)}]{Alekseev:2010mx}%
  \BibitemOpen
  \bibfield  {author} {\bibinfo {author} {\bibfnamefont {G.A.} \bibnamefont
  {Alekseev}}, }in \href {\doibase 10.1142/9789814374552_0033} {\emph {\bibinfo
  {booktitle} {{12th Marcel Grossmann Meeting on General Relativity}}}}
  (\bibinfo {year} {2010}) pp. \bibinfo {pages} {645--666}, \Eprint
  {http://arxiv.org/abs/1011.3846} {arXiv:1011.3846}\BibitemShut {NoStop}%
\bibitem [{\citenamefont {Astorino} and \citenamefont
  {Vigano}(2021)}]{Astorino:2021dju}%
  \BibitemOpen
  \bibfield  {author} {\bibinfo {author} {\bibfnamefont {M.}~\bibnamefont
  {Astorino}} and \bibinfo {author} {\bibfnamefont {A.}~\bibnamefont {Vigano}},
  }\href {\doibase 10.1016/j.physletb.2021.136506} {\bibfield  {journal}
  {\bibinfo  {journal} {\emph {Phys. Lett. B}} }\textbf {\bibinfo {volume}
  {820}}, \bibinfo {pages} {136506} (\bibinfo {year} {2021})}, \Eprint
  {http://arxiv.org/abs/2104.07686} {arXiv:2104.07686}\BibitemShut {NoStop}%
\bibitem [{\citenamefont {Vigan\`o}(2022)}]{Vigano:2022hrg}%
  \BibitemOpen
  \bibfield  {author} {\bibinfo {author} {\bibfnamefont {A.}~\bibnamefont
  {Vigan\`o}}, }\emph {\bibinfo {title} {{Black Holes and Solution Generating
  Techniques}}}, \href@noop {} {Ph.D. thesis}, \bibinfo  {school} {Milan U.}
  (\bibinfo {year} {2022}), \Eprint {http://arxiv.org/abs/2211.00436}
  {arXiv:2211.00436}\BibitemShut {NoStop}%
\bibitem [{\citenamefont {Weyl}(1917)}]{Weyl:1917gp}%
  \BibitemOpen
  \bibfield  {author} {\bibinfo {author} {\bibfnamefont {H.}~\bibnamefont
  {Weyl}}, }\href {\doibase 10.1007/s10714-011-1310-7} {\bibfield  {journal}
  {\bibinfo  {journal} {\emph {Annalen Phys.}} }\textbf {\bibinfo {volume}
  {54}}, \bibinfo {pages} {117} (\bibinfo {year} {1917})}\BibitemShut {NoStop}%
\bibitem [{\citenamefont {Emparan} and \citenamefont
  {Reall}(2002)}]{Emparan:2001wk}%
  \BibitemOpen
  \bibfield  {author} {\bibinfo {author} {\bibfnamefont {R.}~\bibnamefont
  {Emparan}} and \bibinfo {author} {\bibfnamefont {H.S.} \bibnamefont {Reall}},
  }\href {\doibase 10.1103/PhysRevD.65.084025} {\bibfield  {journal} {\bibinfo
  {journal} {\emph {Phys. Rev. D}} }\textbf {\bibinfo {volume} {65}}, \bibinfo
  {pages} {084025} (\bibinfo {year} {2002})}, \Eprint
  {http://arxiv.org/abs/hep-th/0110258} {arXiv:hep-th/0110258}\BibitemShut
  {NoStop}%
\bibitem [{\citenamefont {Belinsky} and \citenamefont
  {Zakharov}(1978)}]{Belinsky:1971nt}%
  \BibitemOpen
  \bibfield  {author} {\bibinfo {author} {\bibfnamefont {V.A.} \bibnamefont
  {Belinsky}} and \bibinfo {author} {\bibfnamefont {V.E.} \bibnamefont
  {Zakharov}}, }\href@noop {} {\bibfield  {journal} {\bibinfo  {journal} {\emph
  {Sov. Phys. JETP}} }\textbf {\bibinfo {volume} {48}}, \bibinfo {pages} {985}
  (\bibinfo {year} {1978})}\BibitemShut {NoStop}%
\bibitem [{\citenamefont {Harmark}(2004)}]{Harmark:2004rm}%
  \BibitemOpen
  \bibfield  {author} {\bibinfo {author} {\bibfnamefont {T.}~\bibnamefont
  {Harmark}}, }\href {\doibase 10.1103/PhysRevD.70.124002} {\bibfield
  {journal} {\bibinfo  {journal} {\emph {Phys. Rev. D}} }\textbf {\bibinfo
  {volume} {70}}, \bibinfo {pages} {124002} (\bibinfo {year} {2004})}, \Eprint
  {http://arxiv.org/abs/hep-th/0408141} {arXiv:hep-th/0408141}\BibitemShut
  {NoStop}%
\bibitem [{\citenamefont {Israel} and \citenamefont
  {Khan}(1964)}]{Israel:1964}%
  \BibitemOpen
  \bibfield  {author} {\bibinfo {author} {\bibfnamefont {W.}~\bibnamefont
  {Israel}} and \bibinfo {author} {\bibfnamefont {A.K.} \bibnamefont {Khan}},
  }\href {\doibase 10.1007/BF02750196} {\bibfield  {journal} {\bibinfo
  {journal} {\emph {Il Nuovo Cimento}} }\textbf {\bibinfo {volume} {33}},
  \bibinfo {pages} {331} (\bibinfo {year} {1964})}\BibitemShut {NoStop}%
\bibitem [{\citenamefont {Komar}(1959)}]{PhysRev.113.934}%
  \BibitemOpen
  \bibfield  {author} {\bibinfo {author} {\bibfnamefont {A.}~\bibnamefont
  {Komar}}, }\href {\doibase 10.1103/PhysRev.113.934} {\bibfield  {journal}
  {\bibinfo  {journal} {\emph {Phys. Rev.}} }\textbf {\bibinfo {volume} {113}},
  \bibinfo {pages} {934} (\bibinfo {year} {1959})}\BibitemShut {NoStop}%
\bibitem [{\citenamefont {Tomimatsu}(1984)}]{10.1143/PTP.72.73}%
  \BibitemOpen
  \bibfield  {author} {\bibinfo {author} {\bibfnamefont {A.}~\bibnamefont
  {Tomimatsu}}, }\href {\doibase 10.1143/PTP.72.73} {\bibfield  {journal}
  {\bibinfo  {journal} {\emph {Progress of Theoretical Physics}} }\textbf
  {\bibinfo {volume} {72}}, \bibinfo {pages} {73} (\bibinfo {year} {1984})},
  \Eprint
  {http://arxiv.org/abs/https://academic.oup.com/ptp/article-pdf/72/1/73/19572990/72-1-73.pdf}
  {https://academic.oup.com/ptp/article-pdf/72/1/73/19572990/72-1-73.pdf}\BibitemShut
  {NoStop}%
\bibitem [{\citenamefont {Bret\'on} \emph {et~al.}(1998)\citenamefont
  {Bret\'on}, \citenamefont {Garc\'{\i}a}, \citenamefont {Manko}, and
  \citenamefont {Denisova}}]{PhysRevD.57.3382}%
  \BibitemOpen
  \bibfield  {author} {\bibinfo {author} {\bibfnamefont {N.}~\bibnamefont
  {Bret\'on}}, \bibinfo {author} {\bibfnamefont {A.A.} \bibnamefont
  {Garc\'{\i}a}}, \bibinfo {author} {\bibfnamefont {V.S.} \bibnamefont
  {Manko}},  and \bibinfo {author} {\bibfnamefont {T.E.} \bibnamefont
  {Denisova}}, }\href {\doibase 10.1103/PhysRevD.57.3382} {\bibfield  {journal}
  {\bibinfo  {journal} {\emph {Phys. Rev. D}} }\textbf {\bibinfo {volume}
  {57}}, \bibinfo {pages} {3382} (\bibinfo {year} {1998})}\BibitemShut
  {NoStop}%
\bibitem [{\citenamefont {Melvin}(1964)}]{Melvin:1963qx}%
  \BibitemOpen
  \bibfield  {author} {\bibinfo {author} {\bibfnamefont {M.A.} \bibnamefont
  {Melvin}}, }\href {\doibase 10.1016/0031-9163(64)90801-7} {\bibfield
  {journal} {\bibinfo  {journal} {\emph {Phys. Lett.}} }\textbf {\bibinfo
  {volume} {8}}, \bibinfo {pages} {65} (\bibinfo {year} {1964})}\BibitemShut
  {NoStop}%
\bibitem [{\citenamefont {Ernst}(1976)}]{Ernst:1976mzr}%
  \BibitemOpen
  \bibfield  {author} {\bibinfo {author} {\bibfnamefont {F.J.} \bibnamefont
  {Ernst}}, }\href {\doibase 10.1063/1.522781} {\bibfield  {journal} {\bibinfo
  {journal} {\emph {J. Math. Phys.}} }\textbf {\bibinfo {volume} {17}},
  \bibinfo {pages} {54} (\bibinfo {year} {1976})}\BibitemShut {NoStop}%
\bibitem [{\citenamefont {Gibbons} \emph {et~al.}(2013)\citenamefont {Gibbons},
  \citenamefont {Mujtaba}, and \citenamefont {Pope}}]{Gibbons:2013yq}%
  \BibitemOpen
  \bibfield  {author} {\bibinfo {author} {\bibfnamefont {G.W.} \bibnamefont
  {Gibbons}}, \bibinfo {author} {\bibfnamefont {A.H.} \bibnamefont {Mujtaba}},
  and \bibinfo {author} {\bibfnamefont {C.N.} \bibnamefont {Pope}}, }\href
  {\doibase 10.1088/0264-9381/30/12/125008} {\bibfield  {journal} {\bibinfo
  {journal} {\emph {Class. Quant. Grav.}} }\textbf {\bibinfo {volume} {30}},
  \bibinfo {pages} {125008} (\bibinfo {year} {2013})}, \Eprint
  {http://arxiv.org/abs/1301.3927} {arXiv:1301.3927}\BibitemShut {NoStop}%
\bibitem [{\citenamefont {Smarr}(1973)}]{Smarr:1973zz}%
  \BibitemOpen
  \bibfield  {author} {\bibinfo {author} {\bibfnamefont {L.}~\bibnamefont
  {Smarr}}, }\href {\doibase 10.1103/PhysRevD.7.289} {\bibfield  {journal}
  {\bibinfo  {journal} {\emph {Phys. Rev. D}} }\textbf {\bibinfo {volume} {7}},
  \bibinfo {pages} {289} (\bibinfo {year} {1973})}\BibitemShut {NoStop}%
\bibitem [{\citenamefont {Gibbons} \emph {et~al.}(2009)\citenamefont {Gibbons},
  \citenamefont {Herdeiro}, and \citenamefont {Rebelo}}]{Gibbons:2009qe}%
  \BibitemOpen
  \bibfield  {author} {\bibinfo {author} {\bibfnamefont {G.W.} \bibnamefont
  {Gibbons}}, \bibinfo {author} {\bibfnamefont {C.A.R.} \bibnamefont
  {Herdeiro}},  and \bibinfo {author} {\bibfnamefont {C.}~\bibnamefont
  {Rebelo}}, }\href {\doibase 10.1103/PhysRevD.80.044014} {\bibfield  {journal}
  {\bibinfo  {journal} {\emph {Phys. Rev. D}} }\textbf {\bibinfo {volume}
  {80}}, \bibinfo {pages} {044014} (\bibinfo {year} {2009})}, \Eprint
  {http://arxiv.org/abs/0906.2768} {arXiv:0906.2768}\BibitemShut {NoStop}%
\bibitem [{\citenamefont {Delgado} \emph {et~al.}(2018)\citenamefont {Delgado},
  \citenamefont {Herdeiro}, and \citenamefont {Radu}}]{Delgado:2018khf}%
  \BibitemOpen
  \bibfield  {author} {\bibinfo {author} {\bibfnamefont {J.F.M.} \bibnamefont
  {Delgado}}, \bibinfo {author} {\bibfnamefont {C.A.R.} \bibnamefont
  {Herdeiro}},  and \bibinfo {author} {\bibfnamefont {E.}~\bibnamefont {Radu}},
  }\href {\doibase 10.1103/PhysRevD.97.124012} {\bibfield  {journal} {\bibinfo
  {journal} {\emph {Phys. Rev. D}} }\textbf {\bibinfo {volume} {97}}, \bibinfo
  {pages} {124012} (\bibinfo {year} {2018})}, \Eprint
  {http://arxiv.org/abs/1804.04910} {arXiv:1804.04910}\BibitemShut {NoStop}%
\bibitem [{\citenamefont {Bardeen} \emph {et~al.}(1973)\citenamefont {Bardeen},
  \citenamefont {Carter}, and \citenamefont {Hawking}}]{Bardeen:1973gs}%
  \BibitemOpen
  \bibfield  {author} {\bibinfo {author} {\bibfnamefont {J.M.} \bibnamefont
  {Bardeen}}, \bibinfo {author} {\bibfnamefont {B.}~\bibnamefont {Carter}},
  and \bibinfo {author} {\bibfnamefont {S.W.} \bibnamefont {Hawking}}, }\href
  {\doibase 10.1007/BF01645742} {\bibfield  {journal} {\bibinfo  {journal}
  {\emph {Commun. Math. Phys.}} }\textbf {\bibinfo {volume} {31}}, \bibinfo
  {pages} {161} (\bibinfo {year} {1973})}\BibitemShut {NoStop}%
\bibitem [{\citenamefont {Abbott} \emph
  {et~al.}(2016{\natexlab{a}})}]{Abbott:2016blz}%
  \BibitemOpen
  \bibfield  {author} {\bibinfo {author} {\bibfnamefont {B.P.} \bibnamefont
  {Abbott}} \emph {et~al.} (\bibinfo {collaboration} {Virgo, LIGO Scientific}),
  }\href {\doibase 10.1103/PhysRevLett.116.061102} {\bibfield  {journal}
  {\bibinfo  {journal} {\emph {Phys. Rev. Lett.}} }\textbf {\bibinfo {volume}
  {116}}, \bibinfo {pages} {061102} (\bibinfo {year} {2016}{\natexlab{a}})},
  \Eprint {http://arxiv.org/abs/1602.03837} {arXiv:1602.03837}\BibitemShut
  {NoStop}%
\bibitem [{\citenamefont {Abbott} \emph
  {et~al.}(2016{\natexlab{b}})}]{Abbott:2016nmj}%
  \BibitemOpen
  \bibfield  {author} {\bibinfo {author} {\bibfnamefont {B.P.} \bibnamefont
  {Abbott}} \emph {et~al.} (\bibinfo {collaboration} {Virgo, LIGO Scientific}),
  }\href {\doibase 10.1103/PhysRevLett.116.241103} {\bibfield  {journal}
  {\bibinfo  {journal} {\emph {Phys. Rev. Lett.}} }\textbf {\bibinfo {volume}
  {116}}, \bibinfo {pages} {241103} (\bibinfo {year} {2016}{\natexlab{b}})},
  \Eprint {http://arxiv.org/abs/1606.04855} {arXiv:1606.04855}\BibitemShut
  {NoStop}%
\bibitem [{\citenamefont {Abbott} \emph
  {et~al.}(2017{\natexlab{a}})}]{Abbott:2017vtc}%
  \BibitemOpen
  \bibfield  {author} {\bibinfo {author} {\bibfnamefont {B.P.} \bibnamefont
  {Abbott}} \emph {et~al.} (\bibinfo {collaboration} {VIRGO, LIGO Scientific}),
  }\href {\doibase 10.1103/PhysRevLett.118.221101} {\bibfield  {journal}
  {\bibinfo  {journal} {\emph {Phys. Rev. Lett.}} }\textbf {\bibinfo {volume}
  {118}}, \bibinfo {pages} {221101} (\bibinfo {year} {2017}{\natexlab{a}})},
  \bibinfo {note} {[Erratum: Phys.Rev.Lett. 121, 129901 (2018)]}, \Eprint
  {http://arxiv.org/abs/1706.01812} {arXiv:1706.01812}\BibitemShut {NoStop}%
\bibitem [{\citenamefont {Abbott} \emph
  {et~al.}(2017{\natexlab{b}})}]{Abbott:2017oio}%
  \BibitemOpen
  \bibfield  {author} {\bibinfo {author} {\bibfnamefont {B.}~\bibnamefont
  {Abbott}} \emph {et~al.} (\bibinfo {collaboration} {LIGO Scientific, Virgo}),
  }\href {\doibase 10.1103/PhysRevLett.119.141101} {\bibfield  {journal}
  {\bibinfo  {journal} {\emph {Phys. Rev. Lett.}} }\textbf {\bibinfo {volume}
  {119}}, \bibinfo {pages} {141101} (\bibinfo {year} {2017}{\natexlab{b}})},
  \Eprint {http://arxiv.org/abs/1709.09660} {arXiv:1709.09660}\BibitemShut
  {NoStop}%
\bibitem [{\citenamefont {Abbott} \emph {et~al.}(2020)}]{Abbott:2020tfl}%
  \BibitemOpen
  \bibfield  {author} {\bibinfo {author} {\bibfnamefont {R.}~\bibnamefont
  {Abbott}} \emph {et~al.} (\bibinfo {collaboration} {LIGO Scientific, Virgo}),
  }\href {\doibase 10.1103/PhysRevLett.125.101102} {\bibfield  {journal}
  {\bibinfo  {journal} {\emph {Phys. Rev. Lett.}} }\textbf {\bibinfo {volume}
  {125}}, \bibinfo {pages} {101102} (\bibinfo {year} {2020})}, \Eprint
  {http://arxiv.org/abs/2009.01075} {arXiv:2009.01075}\BibitemShut {NoStop}%
\bibitem [{\citenamefont {Abbott} \emph
  {et~al.}(2021)}]{LIGOScientific:2021usb}%
  \BibitemOpen
  \bibfield  {author} {\bibinfo {author} {\bibfnamefont {R.}~\bibnamefont
  {Abbott}} \emph {et~al.} (\bibinfo {collaboration} {LIGO Scientific, VIRGO}),
  }\href@noop {} {  (\bibinfo {year} {2021})}, \Eprint
  {http://arxiv.org/abs/2108.01045} {arXiv:2108.01045}\BibitemShut {NoStop}%
\bibitem [{\citenamefont {Poisson} \emph {et~al.}(2011)\citenamefont {Poisson},
  \citenamefont {Pound}, and \citenamefont {Vega}}]{Poisson:2011nh}%
  \BibitemOpen
  \bibfield  {author} {\bibinfo {author} {\bibfnamefont {E.}~\bibnamefont
  {Poisson}}, \bibinfo {author} {\bibfnamefont {A.}~\bibnamefont {Pound}},  and
  \bibinfo {author} {\bibfnamefont {I.}~\bibnamefont {Vega}}, }\href {\doibase
  10.12942/lrr-2011-7} {\bibfield  {journal} {\bibinfo  {journal} {\emph
  {Living Rev. Rel.}} }\textbf {\bibinfo {volume} {14}}, \bibinfo {pages} {7}
  (\bibinfo {year} {2011})}, \Eprint {http://arxiv.org/abs/1102.0529}
  {arXiv:1102.0529}\BibitemShut {NoStop}%
\bibitem [{\citenamefont {Flanagan} and \citenamefont
  {Hughes}(1998)}]{Flanagan:1997sx}%
  \BibitemOpen
  \bibfield  {author} {\bibinfo {author} {\bibfnamefont {E.E.} \bibnamefont
  {Flanagan}} and \bibinfo {author} {\bibfnamefont {S.A.} \bibnamefont
  {Hughes}}, }\href {\doibase 10.1103/PhysRevD.57.4535} {\bibfield  {journal}
  {\bibinfo  {journal} {\emph {Phys. Rev. D}} }\textbf {\bibinfo {volume}
  {57}}, \bibinfo {pages} {4535} (\bibinfo {year} {1998})}, \Eprint
  {http://arxiv.org/abs/gr-qc/9701039} {arXiv:gr-qc/9701039}\BibitemShut
  {NoStop}%
\bibitem [{\citenamefont {Pani} \emph {et~al.}(2015)\citenamefont {Pani},
  \citenamefont {Gualtieri}, \citenamefont {Maselli}, and \citenamefont
  {Ferrari}}]{Pani:2015hfa}%
  \BibitemOpen
  \bibfield  {author} {\bibinfo {author} {\bibfnamefont {P.}~\bibnamefont
  {Pani}}, \bibinfo {author} {\bibfnamefont {L.}~\bibnamefont {Gualtieri}},
  \bibinfo {author} {\bibfnamefont {A.}~\bibnamefont {Maselli}},  and \bibinfo
  {author} {\bibfnamefont {V.}~\bibnamefont {Ferrari}}, }\href {\doibase
  10.1103/PhysRevD.92.024010} {\bibfield  {journal} {\bibinfo  {journal} {\emph
  {Phys. Rev. D}} }\textbf {\bibinfo {volume} {92}}, \bibinfo {pages} {024010}
  (\bibinfo {year} {2015})}, \Eprint {http://arxiv.org/abs/1503.07365}
  {arXiv:1503.07365}\BibitemShut {NoStop}%
\bibitem [{\citenamefont {Kozai}(1962)}]{Kozai:1962zz}%
  \BibitemOpen
  \bibfield  {author} {\bibinfo {author} {\bibfnamefont {Y.}~\bibnamefont
  {Kozai}}, }\href {\doibase 10.1086/108790} {\bibfield  {journal} {\bibinfo
  {journal} {\emph {Astron. J.}} }\textbf {\bibinfo {volume} {67}}, \bibinfo
  {pages} {591} (\bibinfo {year} {1962})}\BibitemShut {NoStop}%
\bibitem [{\citenamefont {Lidov}(1962)}]{LIDOV1962719}%
  \BibitemOpen
  \bibfield  {author} {\bibinfo {author} {\bibfnamefont {M.}~\bibnamefont
  {Lidov}}, }\href {\doibase https://doi.org/10.1016/0032-0633(62)90129-0}
  {\bibfield  {journal} {\bibinfo  {journal} {\emph {Planetary and Space
  Science}} }\textbf {\bibinfo {volume} {9}}, \bibinfo {pages} {719} (\bibinfo
  {year} {1962})}\BibitemShut {NoStop}%
\bibitem [{\citenamefont {Palenzuela} \emph {et~al.}(2014)\citenamefont
  {Palenzuela}, \citenamefont {Barausse}, \citenamefont {Ponce}, and
  \citenamefont {Lehner}}]{Palenzuela:2013hsa}%
  \BibitemOpen
  \bibfield  {author} {\bibinfo {author} {\bibfnamefont {C.}~\bibnamefont
  {Palenzuela}}, \bibinfo {author} {\bibfnamefont {E.}~\bibnamefont
  {Barausse}}, \bibinfo {author} {\bibfnamefont {M.}~\bibnamefont {Ponce}},
  and \bibinfo {author} {\bibfnamefont {L.}~\bibnamefont {Lehner}}, }\href
  {\doibase 10.1103/PhysRevD.89.044024} {\bibfield  {journal} {\bibinfo
  {journal} {\emph {Phys. Rev. D}} }\textbf {\bibinfo {volume} {89}}, \bibinfo
  {pages} {044024} (\bibinfo {year} {2014})}, \Eprint
  {http://arxiv.org/abs/1310.4481} {arXiv:1310.4481}\BibitemShut {NoStop}%
\bibitem [{\citenamefont {Lopes} and \citenamefont
  {Silk}(2014)}]{Lopes:2014dba}%
  \BibitemOpen
  \bibfield  {author} {\bibinfo {author} {\bibfnamefont {I.}~\bibnamefont
  {Lopes}} and \bibinfo {author} {\bibfnamefont {J.}~\bibnamefont {Silk}},
  }\href {\doibase 10.1088/0004-637X/794/1/32} {\bibfield  {journal} {\bibinfo
  {journal} {\emph {Astrophys. J.}} }\textbf {\bibinfo {volume} {794}},
  \bibinfo {pages} {32} (\bibinfo {year} {2014})}, \Eprint
  {http://arxiv.org/abs/1405.0292} {arXiv:1405.0292}\BibitemShut {NoStop}%
\bibitem [{\citenamefont {Brito} \emph {et~al.}(2015)\citenamefont {Brito},
  \citenamefont {Cardoso}, and \citenamefont {Pani}}]{Brito:2015oca}%
  \BibitemOpen
  \bibfield  {author} {\bibinfo {author} {\bibfnamefont {R.}~\bibnamefont
  {Brito}}, \bibinfo {author} {\bibfnamefont {V.}~\bibnamefont {Cardoso}},  and
  \bibinfo {author} {\bibfnamefont {P.}~\bibnamefont {Pani}}, }\href {\doibase
  10.1007/978-3-319-19000-6} {\bibfield  {journal} {\bibinfo  {journal} {\emph
  {Lect. Notes Phys.}} }\textbf {\bibinfo {volume} {906}}, \bibinfo {pages}
  {pp.1} (\bibinfo {year} {2015})}, \Eprint {http://arxiv.org/abs/1501.06570}
  {arXiv:1501.06570}\BibitemShut {NoStop}%
\bibitem [{\citenamefont {Bonga} \emph {et~al.}(2019)\citenamefont {Bonga},
  \citenamefont {Yang}, and \citenamefont {Hughes}}]{Bonga:2019ycj}%
  \BibitemOpen
  \bibfield  {author} {\bibinfo {author} {\bibfnamefont {B.}~\bibnamefont
  {Bonga}}, \bibinfo {author} {\bibfnamefont {H.}~\bibnamefont {Yang}},  and
  \bibinfo {author} {\bibfnamefont {S.A.} \bibnamefont {Hughes}}, }\href
  {\doibase 10.1103/PhysRevLett.123.101103} {\bibfield  {journal} {\bibinfo
  {journal} {\emph {Phys. Rev. Lett.}} }\textbf {\bibinfo {volume} {123}},
  \bibinfo {pages} {101103} (\bibinfo {year} {2019})}, \Eprint
  {http://arxiv.org/abs/1905.00030} {arXiv:1905.00030}\BibitemShut {NoStop}%
\bibitem [{\citenamefont {Cunha} \emph {et~al.}(2018)\citenamefont {Cunha},
  \citenamefont {Herdeiro}, and \citenamefont {Rodriguez}}]{Cunha:2018cof}%
  \BibitemOpen
  \bibfield  {author} {\bibinfo {author} {\bibfnamefont {P.V.P.} \bibnamefont
  {Cunha}}, \bibinfo {author} {\bibfnamefont {C.A.R.} \bibnamefont {Herdeiro}},
   and \bibinfo {author} {\bibfnamefont {M.J.} \bibnamefont {Rodriguez}},
  }\href {\doibase 10.1103/PhysRevD.98.044053} {\bibfield  {journal} {\bibinfo
  {journal} {\emph {Phys. Rev. D}} }\textbf {\bibinfo {volume} {98}}, \bibinfo
  {pages} {044053} (\bibinfo {year} {2018})}, \Eprint
  {http://arxiv.org/abs/1805.03798} {arXiv:1805.03798}\BibitemShut {NoStop}%
\bibitem [{\citenamefont {Papapetrou}(1945)}]{10.2307/20488481}%
  \BibitemOpen
  \bibfield  {author} {\bibinfo {author} {\bibfnamefont {A.}~\bibnamefont
  {Papapetrou}}, }\href {http://www.jstor.org/stable/20488481} {\bibfield
  {journal} {\bibinfo  {journal} {\emph {Proceedings of the Royal Irish
  Academy. Section A: Mathematical and Physical Sciences}} }\textbf {\bibinfo
  {volume} {51}}, \bibinfo {pages} {191} (\bibinfo {year} {1945})}\BibitemShut
  {NoStop}%
\bibitem [{\citenamefont {Majumdar}(1947)}]{PhysRev.72.390}%
  \BibitemOpen
  \bibfield  {author} {\bibinfo {author} {\bibfnamefont {S.D.} \bibnamefont
  {Majumdar}}, }\href {\doibase 10.1103/PhysRev.72.390} {\bibfield  {journal}
  {\bibinfo  {journal} {\emph {Phys. Rev.}} }\textbf {\bibinfo {volume} {72}},
  \bibinfo {pages} {390} (\bibinfo {year} {1947})}\BibitemShut {NoStop}%
\bibitem [{\citenamefont {Herdeiro} and \citenamefont
  {Radu}(2023{\natexlab{a}})}]{Herdeiro:2023mpt}%
  \BibitemOpen
  \bibfield  {author} {\bibinfo {author} {\bibfnamefont {C.A.R.} \bibnamefont
  {Herdeiro}} and \bibinfo {author} {\bibfnamefont {E.}~\bibnamefont {Radu}},
  }\href {\doibase 10.1103/PhysRevD.107.064044} {\bibfield  {journal} {\bibinfo
   {journal} {\emph {Phys. Rev. D}} }\textbf {\bibinfo {volume} {107}},
  \bibinfo {pages} {064044} (\bibinfo {year} {2023}{\natexlab{a}})}, \Eprint
  {http://arxiv.org/abs/2302.00016} {arXiv:2302.00016}\BibitemShut {NoStop}%
\bibitem [{\citenamefont {Herdeiro} and \citenamefont
  {Radu}(2023{\natexlab{b}})}]{Herdeiro:2023roz}%
  \BibitemOpen
  \bibfield  {author} {\bibinfo {author} {\bibfnamefont {C.A.R.} \bibnamefont
  {Herdeiro}} and \bibinfo {author} {\bibfnamefont {E.}~\bibnamefont {Radu}},
  }\href@noop {} {  (\bibinfo {year} {2023}{\natexlab{b}})}, \Eprint
  {http://arxiv.org/abs/2305.15467} {arXiv:2305.15467}\BibitemShut {NoStop}%
\bibitem [{\citenamefont {Emparan}(2000)}]{Emparan:1999au}%
  \BibitemOpen
  \bibfield  {author} {\bibinfo {author} {\bibfnamefont {R.}~\bibnamefont
  {Emparan}}, }\href {\doibase 10.1103/PhysRevD.61.104009} {\bibfield
  {journal} {\bibinfo  {journal} {\emph {Phys. Rev. D}} }\textbf {\bibinfo
  {volume} {61}}, \bibinfo {pages} {104009} (\bibinfo {year} {2000})}, \Eprint
  {http://arxiv.org/abs/hep-th/9906160} {arXiv:hep-th/9906160}\BibitemShut
  {NoStop}%
\bibitem [{\citenamefont {Dias} \emph {et~al.}(2023)\citenamefont {Dias},
  \citenamefont {Gibbons}, \citenamefont {Santos}, and \citenamefont
  {Way}}]{Dias:2023rde}%
  \BibitemOpen
  \bibfield  {author} {\bibinfo {author} {\bibfnamefont {O.J.C.} \bibnamefont
  {Dias}}, \bibinfo {author} {\bibfnamefont {G.W.} \bibnamefont {Gibbons}},
  \bibinfo {author} {\bibfnamefont {J.E.} \bibnamefont {Santos}},  and \bibinfo
  {author} {\bibfnamefont {B.}~\bibnamefont {Way}}, }\href@noop {} {  (\bibinfo
  {year} {2023})}, \Eprint {http://arxiv.org/abs/2303.07361}
  {arXiv:2303.07361}\BibitemShut {NoStop}%
\end{thebibliography}%

\end{document}